\documentclass[a4paper,fleqn,usenatbib,useAMS]{mnras}
\usepackage{color}
\usepackage{amsmath}	
\usepackage{amssymb}	
\usepackage{multicol}        
\usepackage{bm}		

\usepackage{textcomp,amssymb}
\usepackage[pdftex]{epsfig}
\usepackage{leftidx}
\usepackage{hyperref} 

\usepackage{array}
\usepackage{supertabular}
\usepackage{hhline}
\usepackage{multirow}
\usepackage{balance}
\usepackage{times}

\usepackage{bm}

\usepackage[pdftex]{epsfig}
\DeclareGraphicsExtensions{.jpg,.mps,.pdf,.png}

\newcommand{\mP}{{\cal P}}

\newcommand{\vu}{\textbf{u}}

\newcommand{\dd}{\hbox{d}}

\definecolor{darkerred}{rgb}{0.8,0,0}

\begin{document}
\title[Quasi-linear density and velocity statistics]{Two is better than one: joint statistics of density and velocity\\
in concentric spheres as a cosmological probe}

\author[C. Uhlemann, S. Codis, O. Hahn, C. Pichon and F.~Bernardeau]{
\parbox[t]{\textwidth}
{C. Uhlemann$^{1}$, S. Codis$^2$, O. Hahn$^3$, C. Pichon$^{4,5}$ and F.~Bernardeau$^{4,6}$
}
\vspace*{6pt}\\
\noindent
$^{1}$  {Institute for Theoretical Physics, Utrecht University, Princetonplein 5, 3584 CC Utrecht, The Netherlands}\\
$^{2}$ Canadian Institute for Theoretical Astrophysics, University of Toronto, 60 St. George Street, Toronto, ON M5S 3H8, Canada\\
$^{3}$ Laboratoire Lagrange, Universit\'e C\^ote d'Azur, Observatoire de la C\^ote d'Azur, CNRS, Blvd de l'Observatoire, 06304 Nice, France\\
$^{4}$ Sorbonne Universit\'es, UPMC Paris 6 \& CNRS, UMR 7095, Institut d'Astrophysique de Paris,
 98 bis bd Arago, 75014 Paris, France\\
 $^{5}$ Korea Institute of Advanced Studies (KIAS) 85 Hoegiro, Dongdaemun-gu, Seoul, 02455, Republic of Korea\\
$^{6}$ CNRS \& CEA, UMR 3681, Institut de Physique Th\'eorique, F-91191 Gif-sur-Yvette, France\\
}

\maketitle

\begin{abstract}
{The analytical formalism to obtain the probability distribution functions (PDFs) of spherically-averaged cosmic densities and velocity divergences in the mildly non-linear regime is presented. A large-deviation principle is applied to those cosmic fields assuming their most likely dynamics in spheres is set by the spherical collapse model. We validate our analytical results using state-of-the-art dark matter simulations with a phase-space resolved velocity field finding a 2\% percent level agreement for a wide range of velocity divergences and densities in the mildly nonlinear regime ($\sim 10 $Mpc$/h$ at redshift zero), usually inaccessible to perturbation theory. From the joint PDF of densities and velocity divergences measured in two concentric spheres, we extract with the same accuracy velocity profiles and conditional velocity PDF subject to a given over/under-density which are of interest to understand the non-linear evolution of velocity flows.
Both PDFs are used to build a simple but accurate maximum likelihood estimators for the redshift evolution of the variance of both the density and velocity divergence fields, which have smaller relative errors than their sample variances when non-linearities appear. Given  the dependence of the velocity divergence on the growth rate, there is a significant gain in using the full knowledge of both PDFs to derive constraints on the equation of state of dark energy. Thanks to the insensitivity of the velocity divergence to bias, its PDF can be used to obtain unbiased constraints on the growth of structures $(\sigma_8,f)$ or it can be combined with the galaxy density PDF to extract bias parameters.
}
\end{abstract}
\vskip2pc


\maketitle

\section{Introduction}

The kinematics of the large-scale structure of the Universe should allow astronomers to  put very tight constraints on  cosmological models. Deep and wide spectroscopic surveys, like the recently proposed MSE\footnote{\url{http://mse.cfht.hawaii.edu/project}} \citep{2016arXiv160600043M}, will soon allow us to study fundamental physics from the distribution of galaxies, in particular the origin of cosmic acceleration. 
However to reach percent precision on the equation of state of dark energy, astronomers are facing various challenges. In particular, luminous matter is not a fair tracer of the total amount of matter in the Universe \citep{Kaiser84} and is subject to systematic effects like redshift space distortions. Unlike the density field, peculiar velocities are much less affected by galaxy biasing \citep{2011MNRAS.416.1703E,2012ApJ...755...58N,2015PhRvD..92l3507B} and therefore represent a more direct probe of the dynamics. Their power spectrum \citep{2012MNRAS.427L..25J} and skewness \citep{1995MNRAS.274...20B} have therefore been proposed as  unbiased cosmological probes. 

However, in order to extract as much non-linear information as possible, it has become necessary to   investigate alternative estimators to the standard hierarchy of N-point correlation functions of both densities and velocities.
 In this context, \cite{Bernardeau14,Bernardeau15,Codisetal2016}  have shown how the full statistics of cosmic densities in concentric spheres can leverage cosmic parameters 
competitively, as the corresponding spherical symmetry allows for analytical predictions in the mildly non-linear regime, beyond 
what is commonly achievable via standards statistics. Indeed, the zero variance limit  of the cumulant generating functions
yields estimates of the joint probability distribution function (PDF hereafter) which seems to match simulations in the regime of variances of order unity
  \citep{BalianSchaeffer89,Bernardeau92vel,1993ApJ...412L...9J,Valageas02,Bernardeau14,Bernardeau15}.
  Following this work, \cite{Uhlemann16}  recently presented fully analytic expressions for the PDF of  the density field in spheres using a saddle-point approximation applied to the logarithm of the density.
  
This formalism  can  be extended to also include the velocity divergence and obtain the joint statistics of the density and velocity field
while relying on    large deviation statistics and spherical collapse.
This is of interest since  velocities probe more directly the total gravitational field and are more sensitive to the growth rate, unlike e.g. galaxy count-in-cells which are biased with respect to the underlying dark matter density field. The purpose of this paper is therefore to generalize the above-cited  works to the joint statistics of the density field $\rho$ {and} the velocity divergence $\theta$ in multiple concentric spheres. Conditional statistics of velocities in over- and underdense environments can be deduced and are of particular interest in the context of void dynamics. We will also use these statistics to build a maximum likelihood toy model  for the estimation of cosmic parameters. 

The PDFs of the density and peculiar velocity divergence in a sphere have been obtained and compared to simulations first in \cite{BernardeauVdW95} and \cite{Juszkiewicz95} using perturbation theory together with an Edgeworth expansion. \cite{ScoccimarroFrieman96} investigated when loop corrections for unsmoothed fields start to dominate over tree-level contributions (hence signaling a breakdown of perturbation theory), finding that, while this happens for a variance $\sigma\simeq 0.5$ for the density, it does not occur until $\sigma\simeq 1$ for the velocity divergence. The velocity divergence field and its cumulants have been studied for different local collapse models in \cite{ScoccimarroFrieman96,FosalbaGaztanaga98,Ohta03} where the tidal contributions to the reduced cumulants have been calculated and observed to cancel out for smoothed fields, thus rendering the spherical collapse model successful.
 
Here, we refine the theoretical formalism for obtaining the PDF of the velocity divergence field and use state-of-the-art numerical simulations that are able to accurately extract the velocity field. Indeed, while peculiar velocities are notoriously difficult to measure in real data, 
recent progress in estimating the divergence of the velocity \citep{Abel2012,Shandarin2012,Hahn15} in {simulations} now allow us to accurately determine these joint PDF 
and compare them to  analytical predictions.
 Furthermore, we extend earlier theoretical works by \cite{Bernardeau94smoothing} to not only one but multiple concentric spheres and we revisit the non-linearity and stochasticity in the relation of the density and velocity divergence that has been investigated in \cite{BernardeauChodorowski99}.

The outline of the paper is the following. Section~\ref{sec:general_formalism} shortly reviews the general formalism to obtain the probability distribution functions (Section~\ref{sec:PDFconstruction}) for the density and the velocity divergence in concentric spheres based on spherical collapse dynamics and large deviation statistics.
Section~\ref{sec:validation} compares the predictions from the fully analytical saddle point approximation to state-of-the art numerical simulations with a phase-space resolved velocity field.
Section~\ref{sec:applications} presents possible applications of this framework for constraining dark energy and structure formation parameters. It also shortly discusses the potential to infer halo bias from joint density and velocity measurements and the connection to measurements of peculiar velocities which pose the main observational challenge to applying our formalism.
Finally, Section~\ref{sec:conclusion} wraps up and gives an outlook.
Appendix~\ref{sec:code} presents the code we publicly release with this paper while Appendix~\ref{app:PDF} provides a deeper theoretical background to our formalism. Appendix~\ref{app:dispersion} discusses the scatter of the density-velocity relation while the multi-scale cumulants of the density and velocity divergence are derived in Appendix~\ref{app:cumulants}.

\section{Density and velocity in spheres}
This paper aims to predict the one-point statistics of the velocity divergence from first principles using the spherical collapse dynamics in the large-deviation regime following the seminal work of \cite{Bernardeau92vel}. 
In this section, we will first introduce the spherical collapse model which is at the heart of the construction and allows us to connect both the spherically-averaged density and the velocity divergence to the initial density field. {Then, we will shortly explain the algorithm of the simulation which we use to assess the relationship between density and velocity divergence numerically. To assess the validity of the relation between density and velocity divergence in a sphere, we will determine their joint PDF from the simulation and discuss their mean relation and dispersion in the context of the spherical collapse model and perturbation theory.}

\label{sec:general_formalism}

\subsection{Spherical collapse dynamics}
The spherical collapse dynamics relates the final density $\rho$ (normalised to unit mean, hence $\rho=1+\delta$) and velocity divergence $\theta=\nabla\cdot\vu/H_0$, with $H_{0}$ the present-day Hubble constant, in a sphere of size $R$ to the initial density contrast $\tau$ in a sphere of size $r=R\rho^{1/3}$ (mass conservation). This mapping in general cannot be written analytically. Parametric solutions can be found in some regimes but here for simplicity, we rely on a very accurate approximate expression given by
\begin{subequations}
\label{eq:SC}
\begin{align}
\label{eq:sc-density}
\rho_{\rm SC}(\tau)&=\left(1-\frac{\tau}{\nu}\right)^{-\nu}\,,\\
\label{eq:sc-veldiv}
\theta_{\rm SC}(\tau)&= -f(\Omega) \frac{\dd\log\rho(\tau)}{\dd\log\tau}=  -f(\Omega)\tau(1 - \tau/\nu)^{-1}\,,
\end{align}
\end{subequations}
with the growth rate $f(\Omega)=\dd\log D/\dd \log a\approx\Omega^{0.6}$ given in terms of the growth factor $D$ that is related to the matter density parameter $\Omega$.

From equation~\eqref{eq:sc-veldiv} it is clear that the velocity divergence $\theta_{\rm SC}(\tau)$ changes more gently with the initial density $\tau$ than the density $\rho_{\rm SC}(\tau)$, which makes it an interesting target for further investigation. In the linear regime, the velocity divergence is proportional to the density contrast, $\theta=-f(\Omega)\delta$, which is why we will use the rescaled velocity divergence  $\tilde \theta=-\theta/f(\Omega)$ as a variable instead of $\theta$ in the following. 

 The joint spherical collapse mapping from equation~\eqref{eq:SC} can be employed to obtain a relation between the nonlinearly evolved velocity divergence and the density in a sphere giving the relation \citep[see, e.g.][]{BernardeauVdW95,BernardeauChodorowski99,Nadkarni13}
\begin{align}
\label{eq:density-to-veldif}
\tilde \theta_{\rm SC}(\rho) &= \nu\left(\rho^{1/\nu}-1\right)\,,
\end{align}
that has been first given in \cite{Bernardeau92vel} for $\nu=3/2$. Note that since the density is manifestly positive $\rho>0$ we have  $\tilde\theta>\tilde\theta_{\rm min}=-\nu\simeq -1.5$.  In what follows, we will use $\nu=21/13$ instead of $3/2$ because it gives both a better match to the tree-order skewness of the density field and the relation between the density and velocity divergence, see also \cite{Uhlemann16}.

\subsection{Specifications and algorithm of the simulation}
\label{subsec:simulation}
The simulation used in this work was  presented in \cite{Hahn15}: it has a box size $1$ Gpc$/h$, while the number of particles is $1024^3$ particles. The gravitational evolution between redshift $z = 100$ and $z = 0$ has been performed using the tree-PM code L-Gadget 3 from \cite{Angulo12} using a $2048^3$ mesh for the PM force and a force softening of $35\,{\rm kpc}/h$.

When estimating densities, we apply a simple cloud-in-cell (CIC) deposit of all particles to a grid of $1024^3$ cells, followed by a convolution with a spherical top-hat kernel (cf. eq~\ref{eq:top-hat}). We do not deconvolve with the CIC kernel since for our measurements the top-hat is always well resolved at this resolution.

Following \cite{Hahn15},
we measure the velocity field properties in this $N$-body simulation  based on a tessellation of the dark matter sheet \citep{Abel2012,Shandarin2012}. In this approach, since dark matter can be assumed to be perfectly cold, the unperturbed particle lattice in Lagrangian space can be decomposed into a tetrahedral covering. The evolution of the tetrahedra as dictated by the motion of their vertex particles then gives information about the dark matter flow everywhere in space. The single stream density $\rho_s$, e.g., is just the inverse of the volume of each tetrahedron. In multi-stream regions, after collapse, several tetrahedra will overlap a given point in Eulerian space so that the total density is given by the sum over all streams overlapping that point. The mean velocity field, in which we are interested in this paper, is then given by
\begin{equation}
\left< \mathbf{v} \right>(\mathbf{x}) = \frac{\sum_{s\in{\rm S}(\mathbf{x})} \mathbf{v}_s(\mathbf{x})\,\rho_s(\mathbf{x}) }{\sum_{s\in{\rm S}(\mathbf{x})} \rho_s(\mathbf{x})},
\label{eq:vel_average}
\end{equation}
where ${\rm S}(\mathbf{x})$ is the set of all streams $s$ (i.e. tetrahedra) that contain point $\mathbf{x}$ and a subscript $s$ indicates the value of a field at $\mathbf{x}$ on a given sheet. The field $\left<\mathbf{v}\right>$ can be sampled on a regular mesh, of resolution $1024^3$ in our case, for further analysis by calculating all intersections. {The new method for  resolving velocity fields has been shown to remove the shot noise that was present in previous approaches, e.g. when comparing with Fig.~13 in \cite{Hahn15}.}

\begin{figure}
\centering
\includegraphics[width=1.0\columnwidth]{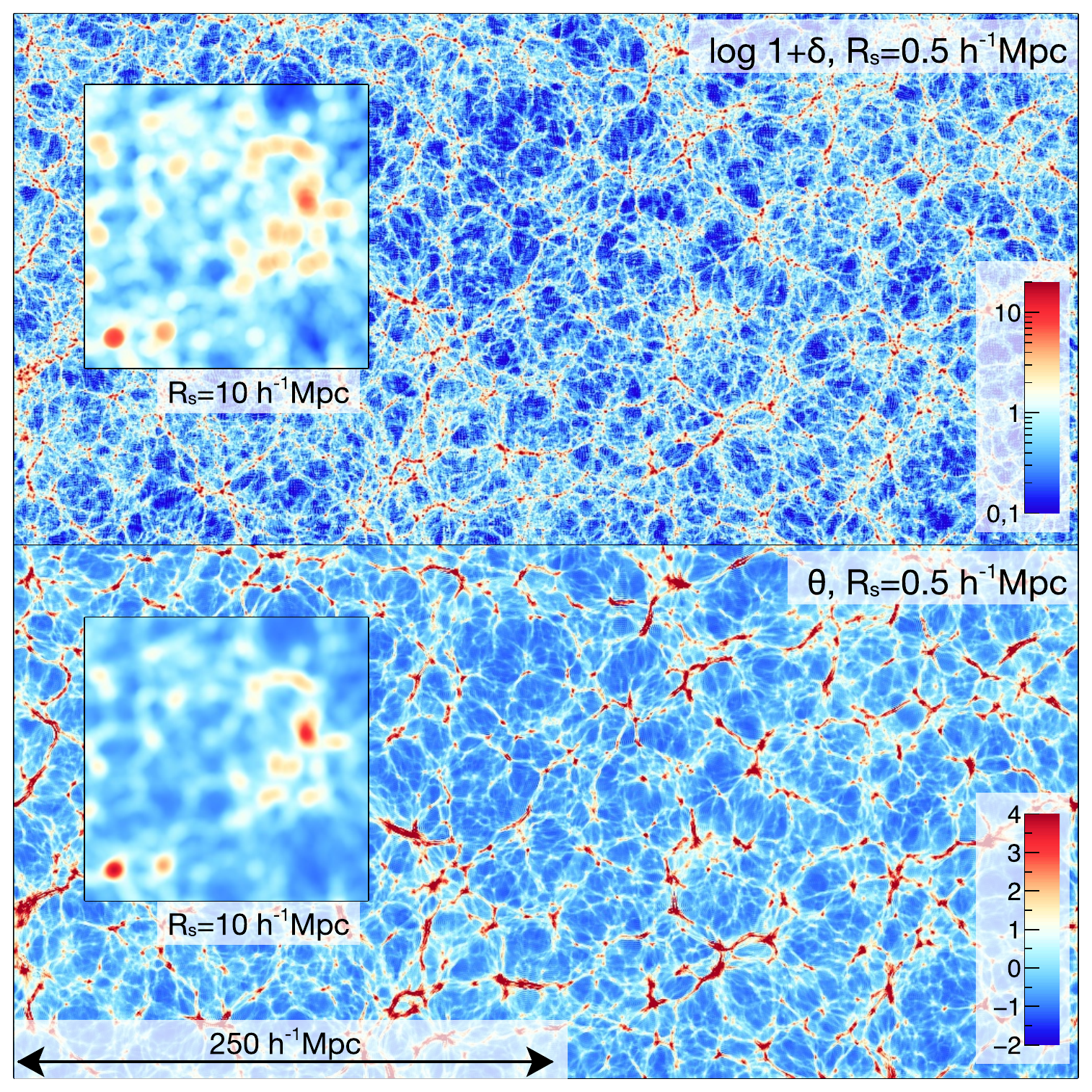}
\caption{Slice of dimension 250$\times$500 $(\text{Mpc}/h)^2$ with a projection depth of 15.6 Mpc$/h$ through the fields of the log-density  (upper panel) and the velocity divergence (lower panel) and the smoothed versions that are under investigation here (insets) of the simulation from \citep{Hahn15}. 
}
   \label{fig:veldiffield}
\end{figure}

In Fig.~\ref{fig:veldiffield}, we show a slice through the simulation for the density field and the velocity divergence. The colour map of the velocity divergence has been inverted with respect to that of the density field to allow for a more easy comparison in terms of the anti-correlation between the two fields in linear theory. For a more detailed description of this approach, we kindly refer   to \cite{Hahn15}. Since we are interested in the velocity divergence smoothed on some scale, we next turn to Fourier transforms to calculate
\begin{equation}
\widetilde{\theta} = i\mathbf{k}\cdot \widetilde{\left<\mathbf{v}\right>}\, \widetilde{W}_{R},
\end{equation}
where a tilde here indicates a Fourier transformed field, and $W_{R}$ is a spherical top-hat in coordinate space of width $R$, see equation~\eqref{eq:top-hat}. 

\subsection{The mean velocity divergence and density relations}
\label{subsec:veldifdensity}
This paper relies on the fact that the velocity divergence on the one hand and the density (both filtered with a spherical top-hat filter) on the 
other hand are closely related to each other, so that the mean trend of their relationship can be captured by constrained averages, that encode the mean velocity divergence  given a certain density and vice versa. Let us  review possible ways to obtain these mean relationships and assess the amplitude of the scatter around this mean, using  a state-of-the-art simulation that resolves the velocity field well.
\begin{figure}
\centering
\includegraphics[width=.95\columnwidth]{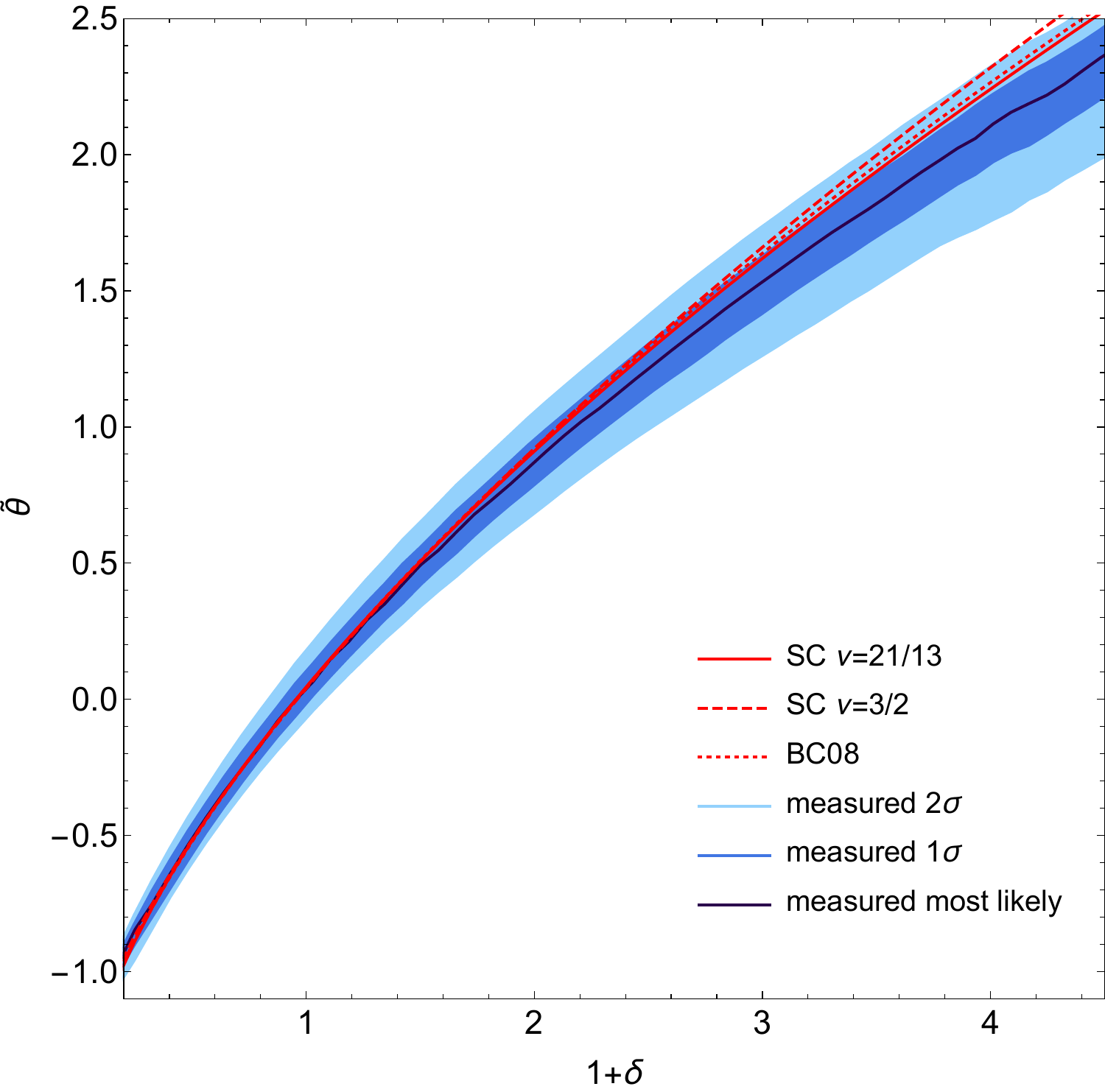}
   \caption{One-sigma (dark blue) and two-sigma (light blue) contours of the conditional PDF of the velocity divergence $\tilde\theta$ given the density $\rho$ in a sphere of identical radius $R=15$ Mpc$/h$ as measured in the simulation. The most likely value of $\tilde\theta$ given $\rho$ is shown with the solid black line. Those measurements are compared to the one-to-one mapping of the spherical collapse dynamics defined in \eqref{eq:density-to-veldif} for different values for the parameter $\nu=3/2$ (dashed red) and $21/13$ (solid red) and to the fit from equation \eqref{eq:Bilicki} (dotted red). Note that for any fitting functions displayed on this figure, we subtract the mean so that $\langle\tilde \theta\rangle$ is always zero by construction. Note also that instead of plotting the most likely value of $\tilde\theta$ given $\rho$, we could have displayed the mean value of $\tilde\theta$ given $\rho$ but the difference is negligible.}
   \label{fig:cond-PDF-rho-theta-sameR}
\end{figure}

In the linear regime the (scaled) velocity divergence is proportional to the density contrast: $\tilde\theta=\delta$. In the nonlinear regime, this linear relationship is modified through the coupling of density and velocity which also induces a scatter in the linear one-to-one relation. As long as the departure from the linear regime is small, generic features can be inferred from perturbation theory \citep{Bernardeau92vel}, which yields an expansion of the mean density given the velocity divergence (and vice versa) in terms of the fields themselves
\begin{align}
\langle\delta\rangle_{\tilde\theta} &= a_1(\sigma_{\tilde\theta}^2) \tilde\theta + a_2(\tilde\theta^2-\sigma_{\tilde\theta}^2) +a_3\tilde\theta^3\,.
\end{align}
To have a consistent expansion up to third order, $a_1=1+\mathcal O(\sigma^2)$ has to be calculated at next-to-leading order, whereas the other coefficients can be computed at their leading order in perturbation theory. The explicit expressions for the coefficients $a_{1,2,3}$ in terms of higher order cumulants of the density and velocity field as computed in \cite{Bernardeau94smoothing} are given in \cite{Bernardeau92vel} and \cite{BernardeauChodorowski99}. More generally, the coefficients $a_n$ were also derived by \cite{ChodorowskiLokas97} and \cite{Chodorowski98} respectively, and their values were explicitly calculated for a Gaussian window function. For a top-hat filter, the values of the higher coefficients $a_{n\geq 2}$ turn out to be independent of the spectral index of the initial power spectrum and to be very close to those obtained from Taylor-expanding equation \eqref{eq:density-to-veldif}.

\cite{Bilicki08} pointed out that deviations from the mean law found from spherical collapse are expected for larger density contrasts and provided the following fitting formula  
\begin{subequations}
\label{eq:Bilicki}
\begin{align}
\tilde \theta_{\rm SC}(\rho) &\stackrel{\rho> 1}{=} 3\left[\rho^{1/2}-\rho^{1/6}\right]\,,\\
\tilde \theta_{\rm SC}(\rho) &\stackrel{\rho < 1}{=} (\rho-1)+(1+\tilde\theta_{\rm min}) \left[\rho\ln\rho - (\rho-1)\right]\,,
\end{align}
\end{subequations}
where the minimum value was parametrised in terms of $\Omega_m$ as $\tilde\theta_{\rm min}=-1-0.5\Omega_m^{0.12-0.06\Omega_m}$. This formula agrees at the percent level with the spherical collapse prediction on equation~\eqref{eq:density-to-veldif} for moderate density contrast $\rho\in[0.1,10]$ which are of interest here, but shows about 10\% deviations for strongly under- or over regions $\rho\in [0.01,0.1]$ or $\rho \in [10,100]$.  The validity of the spherical collapse mapping to connect the velocity divergence and the density is demonstrated in Fig.~\ref{fig:cond-PDF-rho-theta-sameR}. This figure shows that the approximate expression of the spherical collapse given by equation~\eqref{eq:density-to-veldif} provides a very good fit to the mean relation between density and velocity divergence for moderate density contrasts and velocity divergences which is the main focus of this paper. {For a description of the (very slight) dark energy dependence of the density-velocity relation see  \cite{Nadkarni13} where it has also been shown to stay weak even in the non-linear regime.}

\begin{figure}
\includegraphics[width=1\columnwidth]{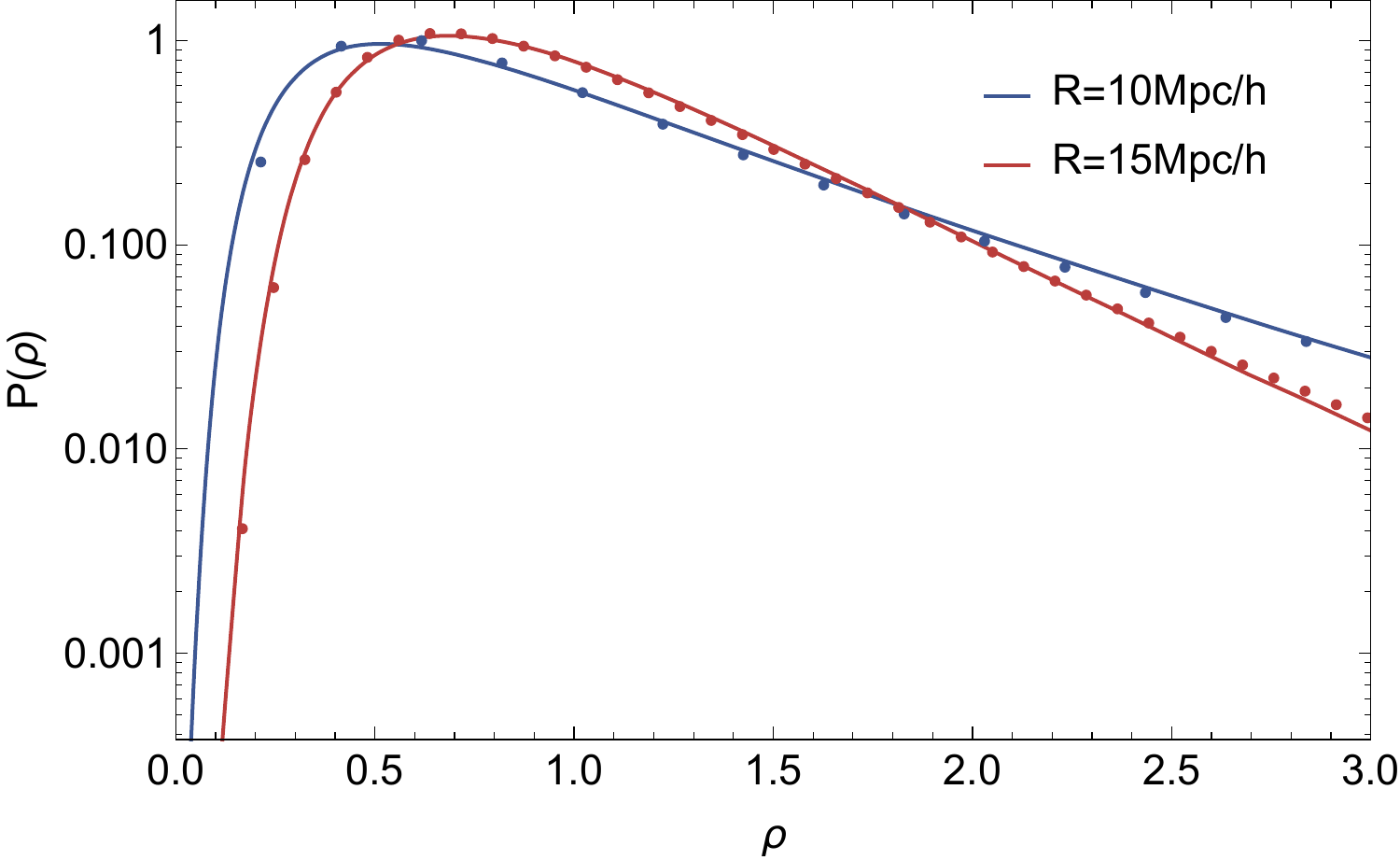}
\caption{Comparison of the density PDF from a direct measurement of the density (dots) and the reconstruction from the normalised velocity divergence $\tilde\theta$ using the spherical collapse mapping equation \eqref{eq:density-to-veldif} with $\nu=21/13$ (solid line) for spheres with radii $R=10$ Mpc$/h$ (blue) and $15$ Mpc$/h$ (yellow). Note that we also subtract the mean in equation \eqref{eq:density-to-veldif}. The  accuracy of the construction in the range of densities from 0.5 to 3 is around 7\%.
}
   \label{fig:PDF-1cell-rho-from-theta}
\end{figure}

 Using this mapping, the PDF of the velocity divergence can be used to reconstruct the PDF of the density (and vice versa) as illustrated in Fig.~\ref{fig:PDF-1cell-rho-from-theta}. Here, we simply map the measured PDF of $\tilde\theta$ to the PDF of the density field using equation~\eqref{eq:density-to-veldif} and subtracting the mean. In practice, it means that we consider the following change of variable
 \begin{eqnarray}
 \label{eq:SCnomean}
 \tilde\theta&\rightarrow&\rho_{\rm SC}(\tilde\theta)=1+\left(1+\frac{\tilde\theta}{\nu}\right)^{\nu}-\Big\langle \!\left(1+\frac{\tilde\theta}{\nu}\right)^{\nu}\!\Big\rangle,\\
 {\cal P}_{\tilde\theta}(\tilde\theta)&\rightarrow& {\cal P}_{\rho}(\rho)=(\rho_{\rm SC}^{-1})'(\rho) {\cal P}_{\tilde\theta}(\rho_{\rm SC}^{-1}(\rho)),
 \end{eqnarray}
 where the mean $\langle (1+{\tilde\theta}/{\nu})^{\nu}\rangle$ is estimated from the measured PDF of the velocity divergence.
  This estimate of the density PDF is successfully compared to the measured PDF of the density field in Fig.~\ref{fig:PDF-1cell-rho-from-theta}, which shows that the spherical collapse dynamics can be used to relate the PDF of the density and velocity divergence field for a broad range of field values and level of non-linearities. The rest of the paper is devoted to derive properly the velocity PDF from first principles using the dynamics of the spherical collapse in the large deviation regime, not only for one but also for two concentric spheres.
 
For a discussion of the dispersion around the mean relation between velocity divergence and density, we refer the reader to Appendix~\ref{app:dispersion}. 

\section{Construction of the PDFs}
\label{sec:PDFconstruction}

Extending upon \cite{Bernardeau94smoothing} and  \cite{Uhlemann16}, 
let us now present the general formalism of how the joint PDF of densities and velocity divergences in concentric spheres can be obtained using large deviation statistics and spherical collapse dynamics, relying on a logarithmic remapping together with a saddle approximation. As described in Section~\ref{sec:general_formalism}, the spherical collapse model provides us with a direct mapping between densities and velocity divergences that will be used to determine the PDF of the velocity divergence using the predictions for the density PDF.

\subsection{Large deviation statistics}

We are interested in the joint statistics of the density $\rho$ and the velocity divergence $\theta$ in $N$ concentric spherical cells of radii $\{R_{k}\}$, indexed by $1\leq k\leq N$ in which the mean density and velocity divergence are denoted by $\Phi=\{\Phi_{k}\}_{1\leq k\leq N}=\{(\rho_{k},\tilde\theta_{k})\}_{1\leq k\leq N}$.

In the following we introduce the essential ingredients that are needed to construct the PDFs of densities and velocities in spheres from initially Gaussian statistics. In the main text we limit ourselves to present the result obtained for the density field \citep{Uhlemann16} and use it to obtain a formula for the velocity divergence using the mean relation discussed in Section~\ref{subsec:veldifdensity}. See Appendix~\ref{app:PDF} for a more general and comprehensive description of the three basic steps for large deviation statistics that allow us to map the initially Gaussian statistics to the final ones. For more details or mathematical rigour we refer to \cite{Bernardeau14,Bernardeau15,LDPinLSS,Uhlemann16}.

\paragraph*{Initial conditions}
 The initial decay-rate function $\Psi(\{\tau_k\}) $ describes the exponential decay of the PDF of the initial density contrasts $\{\tau_k\}$. We assume Gaussian initial conditions which makes the decay-rate function a quadratic form
\begin{equation}
\Psi_i(\{\tau_k\}) = \frac{1}{2} \sum_{i,j} \Xi_{ij}(\{r_k\}) \tau_i\tau_j\,,
\end{equation}
where the precision matrix, $\Xi_{ij}$, is the inverse of the cross-correlation (or covariance) matrix $\Sigma_{ij}=\langle \tau_i\tau_j\rangle$.

\paragraph*{Spherical collapse map from initial to final densities}
To predict the statistics of spherically-averaged quantities one can take advantage of spherical collapse dynamics. This idea is rooted in the so-called contraction principle from large deviation statistics saying that any large deviation follows the least unlikely of all unlikely transformations between the initial and final state because the PDF is exponentially suppressed when departing from the most likely way. Hence, one can use the one-to-one mapping between the initial density contrast $\tau_k$ in a sphere of radius $r_k$ and the final density $\rho_k=\rho_{\rm SC}(\tau_k)$ within radius $R_k$ (with $r_k=\rho_k^{1/3} R_k$ due to mass conservation) to determine the final decay-rate function as
\begin{align}
\Psi_{f}(\{\rho_k\})=\Psi_{i}\left(\{\tau_k\!:\!\rho_k=\rho_{\rm SC}(\tau_k)\}\right).
\end{align}
Note that, while mapping the decay-rate functions with spherical collapse is allowed, one cannot simply in practice map the PDFs with spherical collapse because the relation between initial and final radii arising from mass conservation mixes scales such that fixing a final radius is not identical to fixing the initial radius. To avoid the problem of directly mapping PDFs,  this decay-rate function can be used to derive the cumulant generating function and from there reconstruct the PDF. One can find a saddle point approximation, applied to a suitably mapped set of density variables $\mu_j(\{\rho_k\})$ to implement this procedure efficiently  
 \begin{align}
 \label{eq:saddlePDFrhoN-cell}
\mP(\{\rho_k\})&= \left|\det\left[\frac{\partial\mu_{i}}{\partial \rho_{j}}\right]\right| \sqrt{\det\left[\frac{\partial^{2}\Psi_f}{\partial \mu_{i}\partial \mu_{j}}\right]} \frac{ \exp\left[-\Psi_f\right]}{ (2 \pi)^{N/2}} 
\,,
 \end{align}
\paragraph*{Spherical collapse map from densities to velocity divergences}
Thanks to the one to one mapping between final density and velocity divergence, as shown in more detail in Appendix~\ref{app:PDF}, one can simply rely on the results obtained for the density PDFs in \cite{Uhlemann16} to obtain the PDFs of the velocity divergence and joint PDFs by a simple remapping
 \begin{align}
 \label{eq:saddlePDFrhothetaN-cell}
\mP(\{\rho_k\},\{\tilde\theta_{\tilde k}\})&= \mP\left(\{\rho_k\},\{\rho_{\tilde k}=\rho_{\rm SC}(\tilde\theta_{\tilde k})\}\right) \prod_{\tilde k} \rho'(\tilde\theta_{\tilde k})
\,,
 \end{align}
In the main text, we only discuss the validity of the prediction for the PDF. For a discussion of the accuracy of the tree-order perturbation theory prediction for cumulants of the velocity divergence and density field, we refer the reader to Appendix~\ref{app:cumulants}. Note that the assumption that the mapping between the density and velocity divergence is one-to-one restricts the application of these formulas to cases where the radii $R_k$ and $R_{\tilde k}$ are sufficiently different from each other, since in the limit $R_{\tilde k}\rightarrow R_k$ one would obtain a delta function centred around the spherical collapse relation which is not realistic as shown in Fig.~\ref{fig:joint-PDF-theta-scatterrho-sameR}. In practice, our predictions work reasonably well for radii of $R_1=10$ and $R_2=15$ Mpc$/h$ while the resolution of the simulation does not allow to make a definite statement for the joint PDF with $R_1=10$ and $R_2=11$ Mpc$/h$.
\paragraph*{Ensuring the correct normalization and mean}
\label{sec:norm}
Due to the remappings, the saddle point PDFs have to be modified to ensure normalization and the correct mean and variance for both the density and the velocity divergence. Let us illustrate this procedure based on the one cell case where it can be easily implemented using equation~\eqref{eq:saddlePDFthetafromlogrhonorm}.

Since the saddle-point method yields only an approximation to the exact PDF, the PDF obtained from equation\,\eqref{eq:saddlePDFrhoN-cell} has to be normalised  $\hat\mP_R(\rho) = \mP_R(\rho)/{\textstyle \int \dd\rho}\, \mP_R(\rho)$. Then, since we apply the saddle point approximation to a transformed variable we have to impose the correct mean of this variable in order to ensure  the correct mean of the density $\langle\rho\rangle=1$ and velocity divergence $\langle\theta\rangle=0$. The formulas presented for the PDF so far assumed zero mean of the transformed variable and hence for the log-mapping $\tilde\rho=\exp\mu$. If we let the mean $\bar \mu$ unconstrained, we get instead
$\rho=\exp(\mu-\bar \mu)$. From this equation, one can derive the PDF of the physical density $\rho=\tilde\rho/\exp\bar\mu$
\begin{equation}
{\cal P}_{\rho}(\rho)=\frac{\exp \bar \mu}{\rho} {\cal P}_{\mu}\left(\mu=\log \tilde\rho=\log(\rho\cdot\exp\bar\mu)\right)\,.
\end{equation}
In this equation, one can either predict $\bar \mu$ using perturbation theory, measure it from -- or fit it to -- data, or impose the mean density.

For the density, the mean density should be set to one which requires to determine
\begin{equation}
\exp\bar \mu= \langle\tilde\rho\rangle= \int \tilde\rho\ \mP_\mu(\mu=\log\tilde\rho) \ \dd \tilde\rho \,.
\end{equation}
and use this as normalization for the mean in the PDF
\begin{align}
\label{eq:saddlePDFthetafromlogrhonorm}
\hat\mP_{R}(\rho) &= \mP_{R}\left(\rho \cdot \langle\tilde\rho\rangle\right)  \cdot \langle\tilde\rho\rangle\,.
\end{align}
In addition, for the velocity divergence we have to enforce zero mean, $\langle \tilde\theta\rangle=0$, according to equation~(\ref{eq:SCnomean}).
Similarly, one can  determine the correct normalisation of the mean for the two-cell PDFs of the density or velocity divergence by imposing the means. 
\begin{figure*}
\includegraphics[width=1\columnwidth]{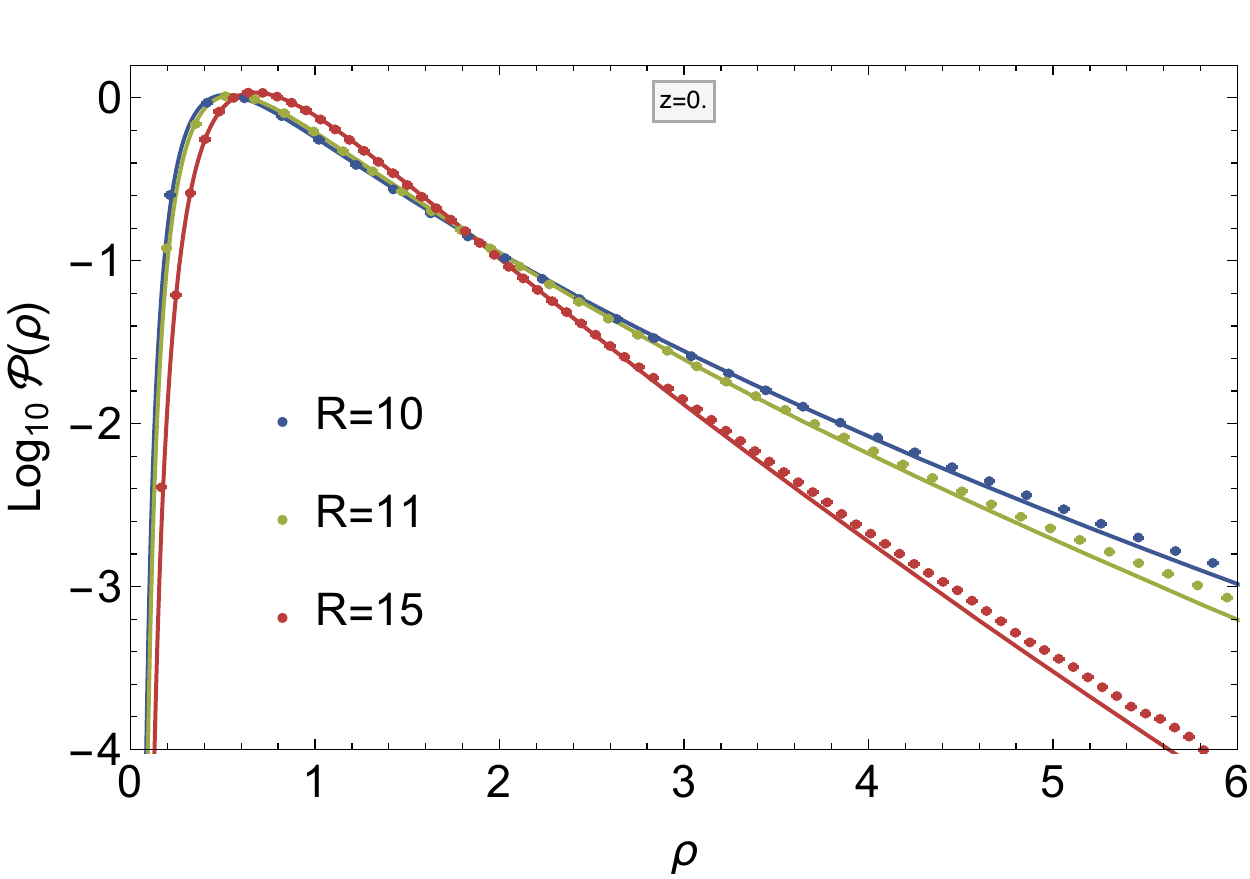}\quad\includegraphics[width=1\columnwidth]{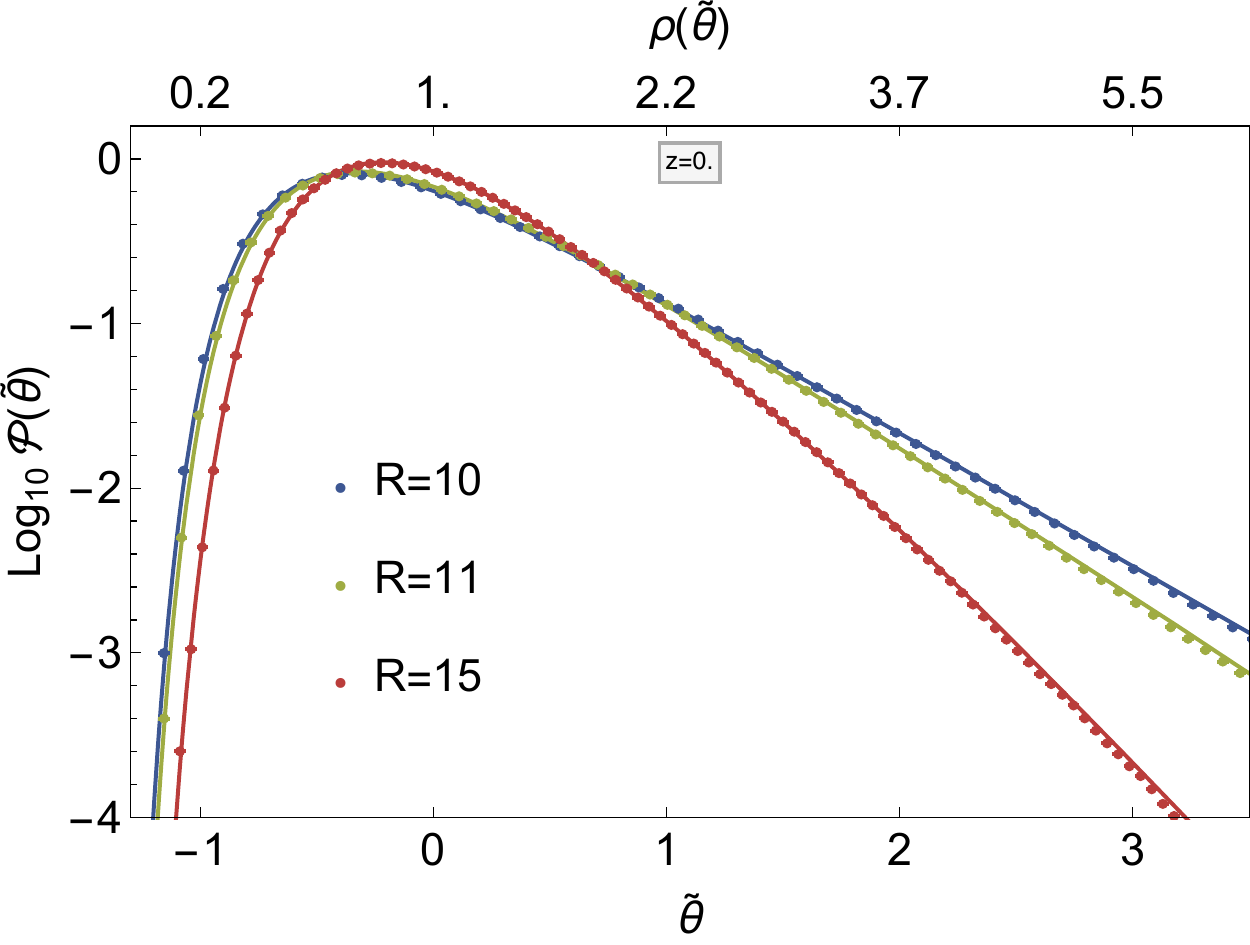}\\
\includegraphics[width=1\columnwidth]{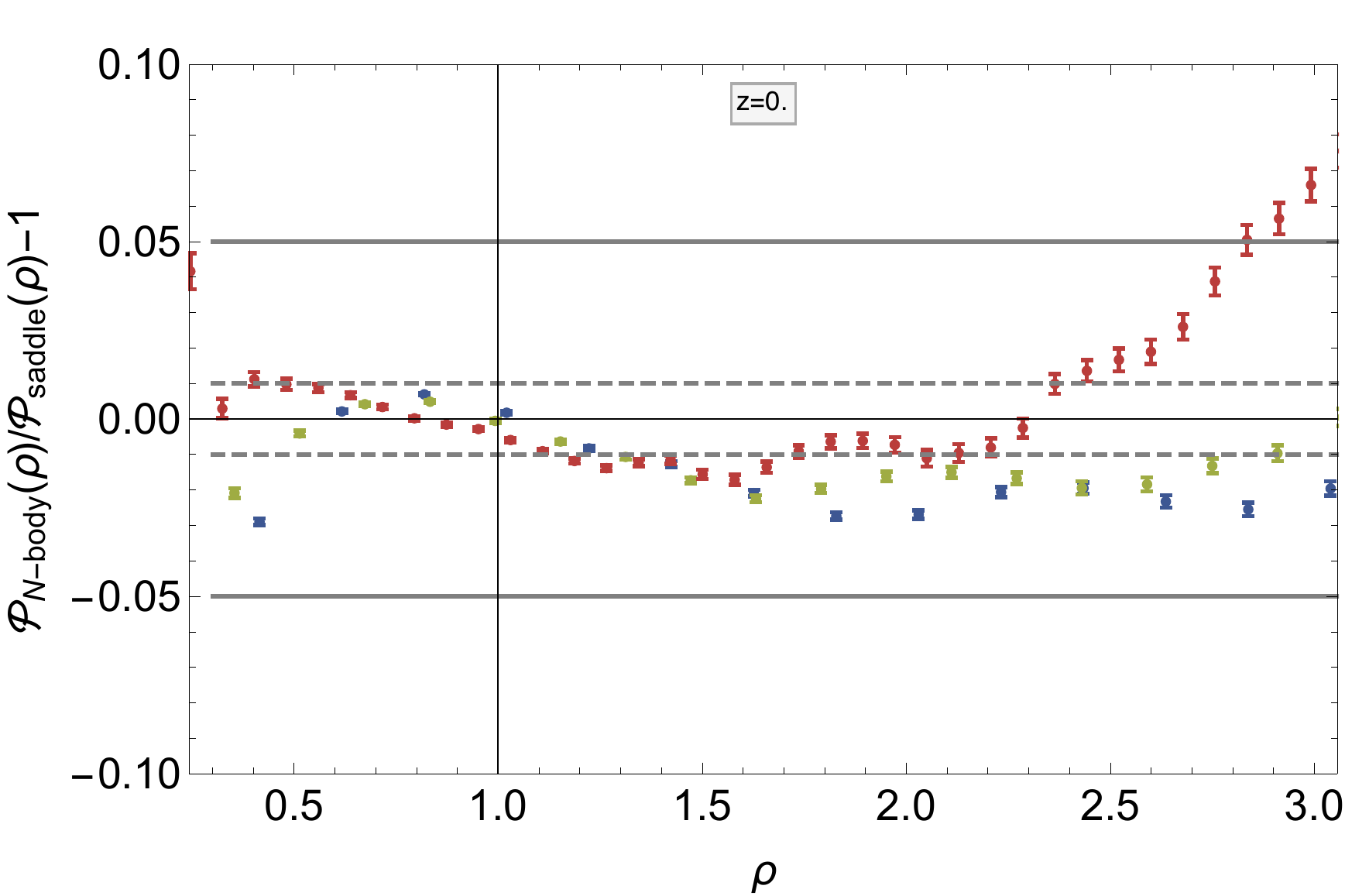}\quad \includegraphics[width=1\columnwidth]{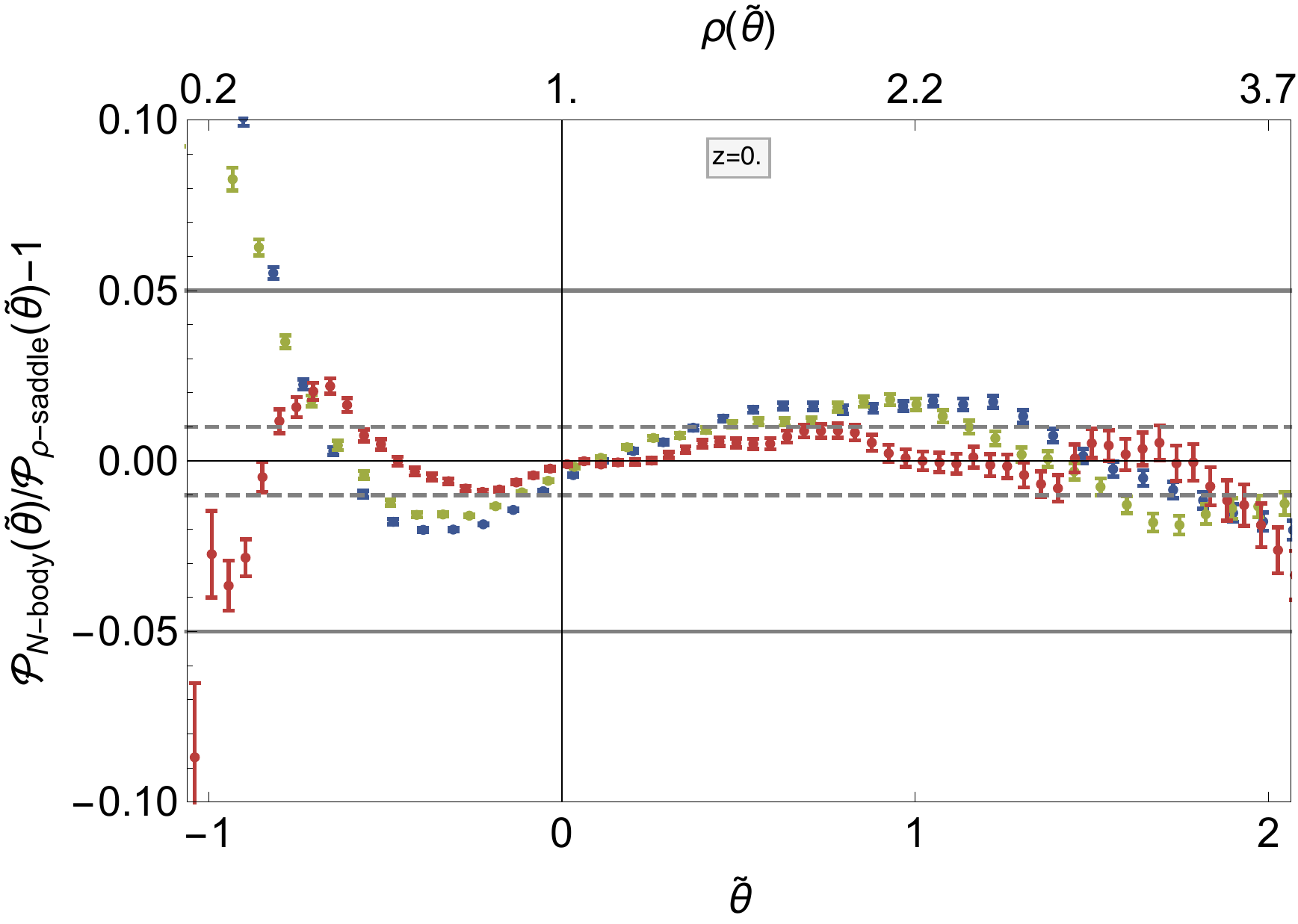}
\caption{Comparison of the PDF of the density $\rho$ {\it (left panel)} and reduced velocity divergence $\tilde\theta$ {\it (right panel)} for radii $R=10,11,15$ Mpc$/h$ {\it (upper panel)} measured in the $N$-body simulation (points) with the saddle point approximations for the log-density and the corresponding residuals for the measured variance (lower panel). 
Note that, as expected, the $\theta$ PDF is well matched over a wider range of equivalent $\rho$.}
   \label{fig:PDF-1cell-rhotheta-num-vs-theo}
\end{figure*}

\subsection{Parametrizing the covariance}
In practice, in order to determine the decay-rate function and therefore the PDFs, one needs to compute the covariance matrix between initial densities in spheres of radii $r_{i}$ and $r_{j}$
\begin{equation}
\sigma^2(r_i,r_j) = \int\frac{\dd^{3} k}{(2\pi)^{3}}P^{\rm lin}(k)W_{\rm 3D}(k r_i)W_{\rm 3D}(k r_j)\,,
\label{eq:defSigma2}
\end{equation}
where  $W_{\rm 3D}(k)$ is the shape of the top-hat window function in Fourier space,
 \begin{equation}
 \label{eq:top-hat}
W_{\rm 3D}(k)=3\sqrt{\frac{\pi}{2}}\frac{J_{3/2}(k)}{k^{3/2}}\,,
\end{equation}
and $J_{3/2}(k)$ the Bessel function of the first kind of order $3/2$. Given an initial power spectrum, we can tabulate the full expression for the variance $\sigma^2(R)$ and the covariance for two distinct radii $\sigma^2(R_i,R_j)$ from the initial power spectrum by relying on  \href{http://cita.utoronto.ca/~codis/LSSFast.html}{LSSFast} \citep{Codisetal2016}. The range of initial radii $r$ that needs to be considered in the tabulation  depends on the final radius and densities in the corresponding sphere according to mass conservation $r=\rho^{1/3}R$. For radii of $R\simeq 10-15$ Mpc$/h$ and densities in the range $\rho\in[0.1,5]$, we typically need to cover $r\in[5,25]$ Mpc$/h$.
Alternatively, to obtain straightforward analytical expressions, it can be useful to parametrise the covariance matrix for a power-law linear power spectrum with spectral index $n$ by 
\begin{subequations}
\label{eq:sigijparam}
\begin{align}
\sigma^{2}(r_{i},r_{i})&=\sigma^2(R_{p})\left(\frac{r_{i}}{R_{p}}\right)^{-n(R_{p})-3}\,,\\
\sigma^{2}(r_{i},r_{j> i})
&=\sigma^2(R_{p})\,{\cal G}\left(\frac{r_i}{R_p},\frac{r_j}{R_p},n(R_{p})\right)\,,
\end{align}
\end{subequations}
where
\begin{align}
{\cal G}(x,y,n)&=\frac{\displaystyle \int{\dd^{3}k\,}k^{n}W_{\rm 3D}(k x)W_{\rm 3D}(k y)}{\displaystyle\int{\dd^{3}k\,}k^{n}W_{\rm 3D}(k R_{p})W_{\rm 3D}(k R_{p})}\nonumber
\\
&=
  \! \frac{ (x\!+\!y)^{\alpha} \!\! \left(\!x^2\!+\!y^2\!-\!\alpha x y\right)\!-\!(y\!-\!x)^{\alpha} 
 \!  \!\left(\!x^2\!+\!y^2\!+\!\alpha x y\right)}
   {2^{\alpha}(n+1) x^3 y^3  },\nonumber
\end{align}
with $\alpha=1-n$. 
In what follows, we will use the exact linear power spectrum of the simulation to predict one-cell PDFs but the parametric expressions for the two-cell case, while we have checked that the running of the power spectrum has no significant impact on the two-cell results.

The key parameter in the prediction of the PDF is the value of the variance at the pivot scale. In practice, we will treat the variance of the log density at the pivot scale, $\sigma^2(R_p)$, as measured from or adjusted to the simulations as its linear prediction is not sufficient to get accurate predictions. 
Alternatively, the non-linear value of this parameter could be modeled e.g. following \cite{ReppSzapudi16}\footnote{We find that, while the prediction from linear theory systematically overestimates the nonlinear variance, the lognormal model 
$\sigma_{\mu}^2 =\log \left(1+{\sigma_{\rm lin}^2}\right)$
is much closer to the measured variances probed here.}.
We report the results for the measured variance of  the density $\rho$, velocity divergence $\theta$, the log-density $\mu=\log\rho$ and the equivalent logarithm of the density-analogue for the velocity divergence $\mu_\theta=\nu\log(1+\theta/\nu)$ in Table~\ref{tab:variance}.
      
\begin{table}
\centering
\begin{tabular}{c||c|c|c|c|c}
$R$ [Mpc/$h$] & 5 & 7 & 10 & 11& 15 \\\hline\hline
$\sigma_{\rm lin}$ & 1.060  & 0.865 & 0.683 & 0.638 & 0.506 \\
$\sigma_{\mu,\rm{lognorm}}$ & 0.868 & 0.747 & 0.618 & 0.584 & 0.477 \\\hline
$\hat\sigma_\rho$ & 1.339 &  1.012 & 0.751 & 0.689 & 0.529 \\
$ \hat\sigma_\mu$ & 0.835 & 0.735 & 0.619 & 0.585 & 0.481\\\hline
$\hat\sigma_{\tilde\theta}$ & 0.947 & 0.786 & 0.631 & 0.592 & 0.475 \\
$\hat\sigma_{\mu_{\tilde\theta}}$ & 0.810 & 0.704 & 0.585 & 0.556 & 0.459\\
\end{tabular}
\caption{Variances of the density $\rho$, normalized velocity divergence $\tilde\theta$, the log-density $\mu=\log\rho$ and the equivalent $\mu_{\tilde\theta}=\nu\log(1+\theta/\nu)$ for different radii $R$ at redshift $z=0$ as predicted from linear theory, the lognormal model and measured from the simulation.}
\label{tab:variance}
\end{table}


\section{Validation of the PDFs against  simulations}
\label{sec:validation}

We will now validate our predictions for the one-cell PDF of the density and velocity divergence and the joint PDF of density and velocity divergence at two different scales using the simulation which relies on phase-space resolved velocity field. {Thanks to improvements in both the theoretical formalism and the numerical simulation, we will present a  precise comparison of residuals for the one-cell PDF of the velocity divergence together with first results on the joint PDF of densities and velocity divergences measured at two different scales.}

\subsection{One cell PDF comparison}

Fig.~\ref{fig:PDF-1cell-rhotheta-num-vs-theo} shows a comparison of the measurements in the $N$-body simulation described in \ref{subsec:simulation} at redshift $z=0$ for radii $R=10,11,15$ Mpc$/h$ and the analytical saddle point approximation for the PDF of the density and velocity divergence, respectively. The measured variances are given in Table~\ref{tab:variance}.

For the density PDF, as already found in \cite{Uhlemann16} and \cite{Uhlemann16b}, the agreement is very good, below the 2\% level for all densities from 0.5 to 2.5 for all radii considered here. This result is impressing given that it is obtained at lower redshift and expected to improve significantly for higher redshifts where the variances (which are about 0.5 here) are reduced. We can also see that the predicted rare event tails agree qualitatively very well with the simulation results both at the large density and low-density ends.

\begin{figure}
\includegraphics[width=1\columnwidth]{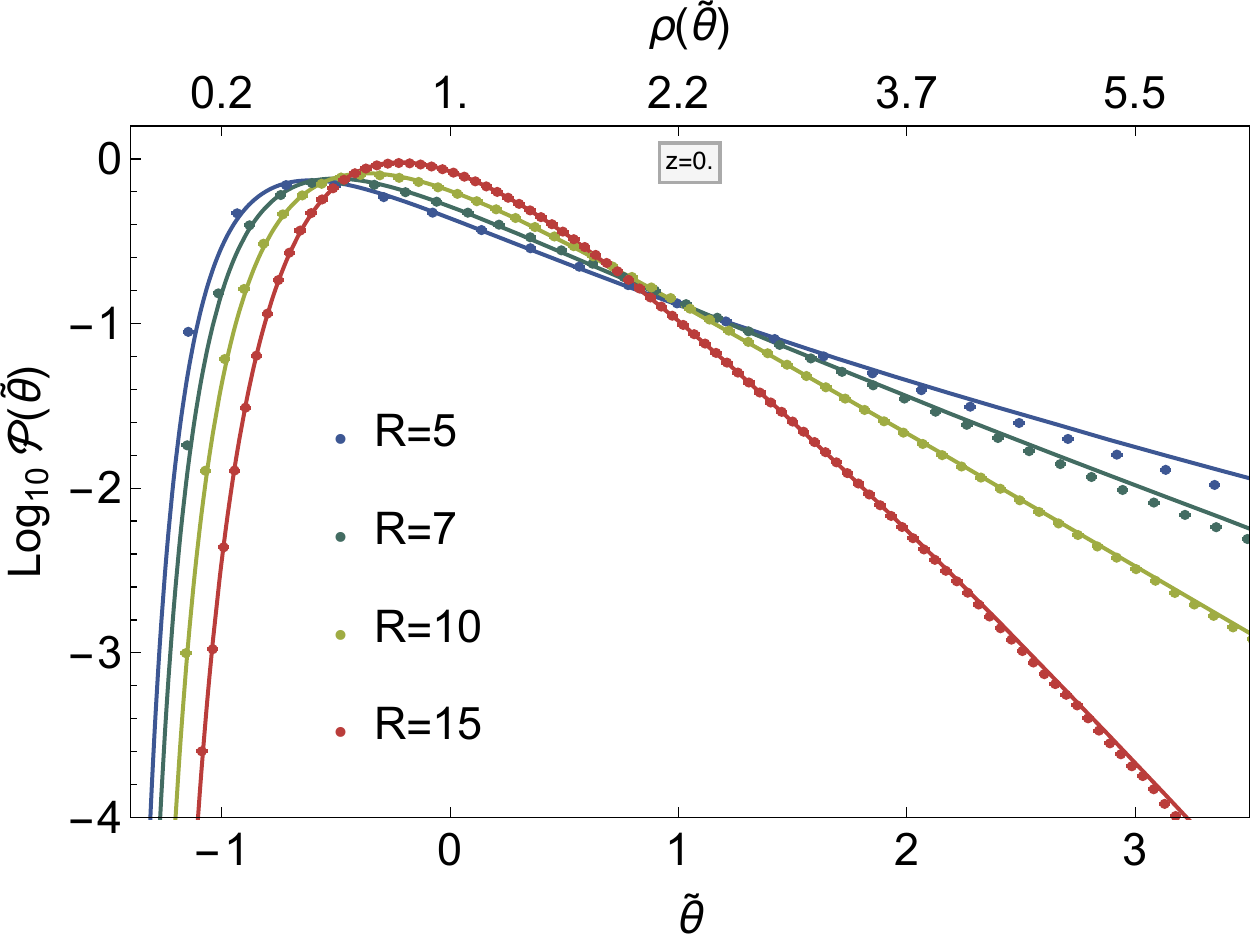}
\includegraphics[width=1\columnwidth]{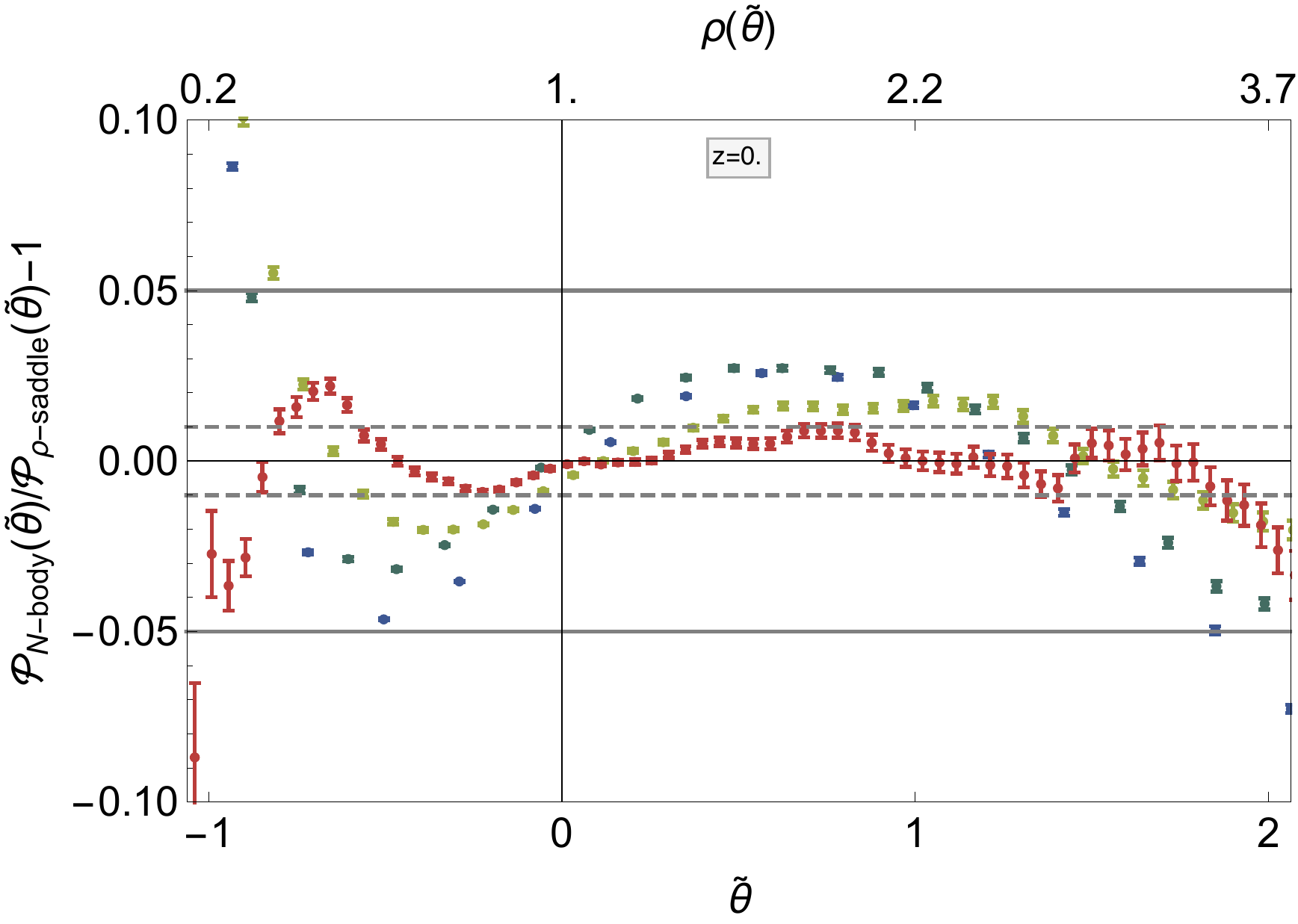}
\caption{Comparison of the velocity divergence PDF as shown in the right panel of Fig.~\ref{fig:PDF-1cell-rhotheta-num-vs-theo} but down to smaller radii $R=5,7,10,15$ Mpc$/h$ {\it (upper panel)} as measured in the $N$-body simulation (points) with the saddle point approximations for the log-density and the corresponding residuals for the measured variance (lower panel).}
   \label{fig:PDF-1cell-rhotheta-num-vs-theo-smallR}
\end{figure}

For the PDF of the velocity divergence,  first  note the significant improvement on the error bars of the simulation under consideration here compared to the results obtained in \cite{BernardeauVdW95} from a Voronoi and a Delaunay tesselations. This improvement in the accuracy of the velocity measurements allows us to much better test the theoretical predictions from large deviation statistics. Similar to the density field, we  find very good agreement between the analytical prediction and the simulation, at the 2\% level for reduced velocity divergences from $-0.8$ to $2$ that correspond to densities between $0.5$ and $3.5$. Also the rare event tails of the distribution are well-described by the analytical formula with a good match for both large positive velocity divergences and strongly negative velocity divergences. Because of the very good agreement between the prediction and the measurements, we show the same result in Fig.~\ref{fig:PDF-1cell-rhotheta-num-vs-theo-smallR} but extended to smaller radii $R=5$ and $7$ Mpc$/h$ finding agreement at the 5\% level even for those significantly smaller and hence more nonlinear scales where the variances become of order 1. This shows that the PDF of the velocity divergence is even more robust against changes in the variance than the density, which allows to probe smaller scales. Appendix~\ref{app:cumulants} and especially Figs.~\ref{fig:cumulants}~and~\ref{fig:cumulants-log} show how this property also manifests on the reduced cumulants which remain close to the tree-order predictions obtained from perturbation theory at small radii where predictions for the density field deviate. This property of the velocity field was already pointed out in the early works of \citep{1994ApJ...420...44K,ScoccimarroFrieman96twopt,FosalbaGaztanaga98}, in particular in \cite{ScoccimarroFrieman96} where it was explicitly stated that for the velocity divergence higher order corrections only start dominating over the tree-order result when the variance approaches unity.

\subsection{Two cell and conditional PDF comparison}
Let us now present the first results for the joint PDFs of velocity divergences measured at different smoothing scales, the joint PDFs of the velocity divergence and the density measured at different smoothing scales and demonstrate how the joint PDF can be used to obtain predictions for the constrained PDF of the velocity divergence given an over- or under-dense environment.

\subsubsection{The joint PDFs of velocity divergence and density}

\begin{figure}
\centering
\includegraphics[width=0.8\columnwidth]{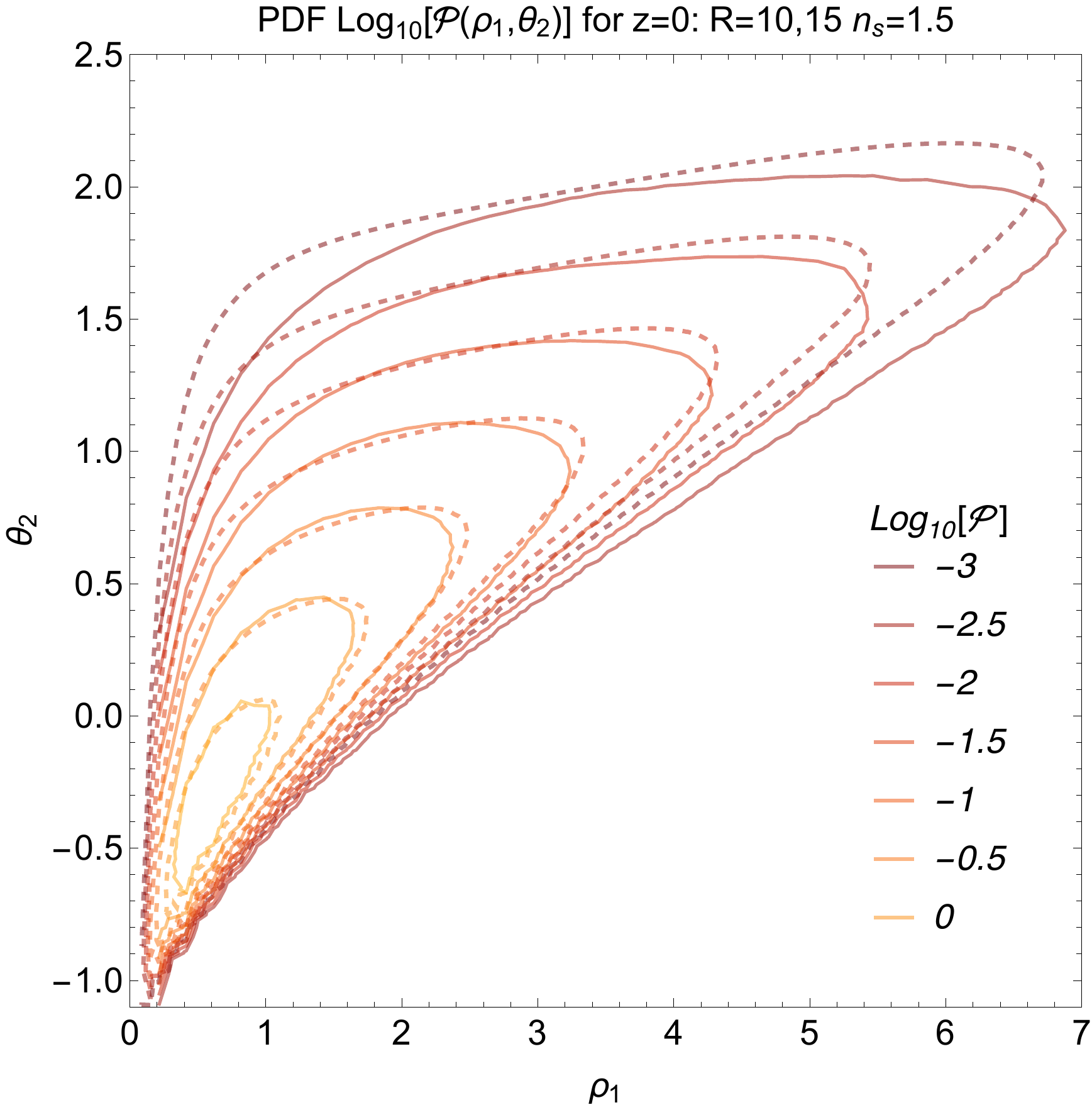}
\includegraphics[width=0.8\columnwidth]{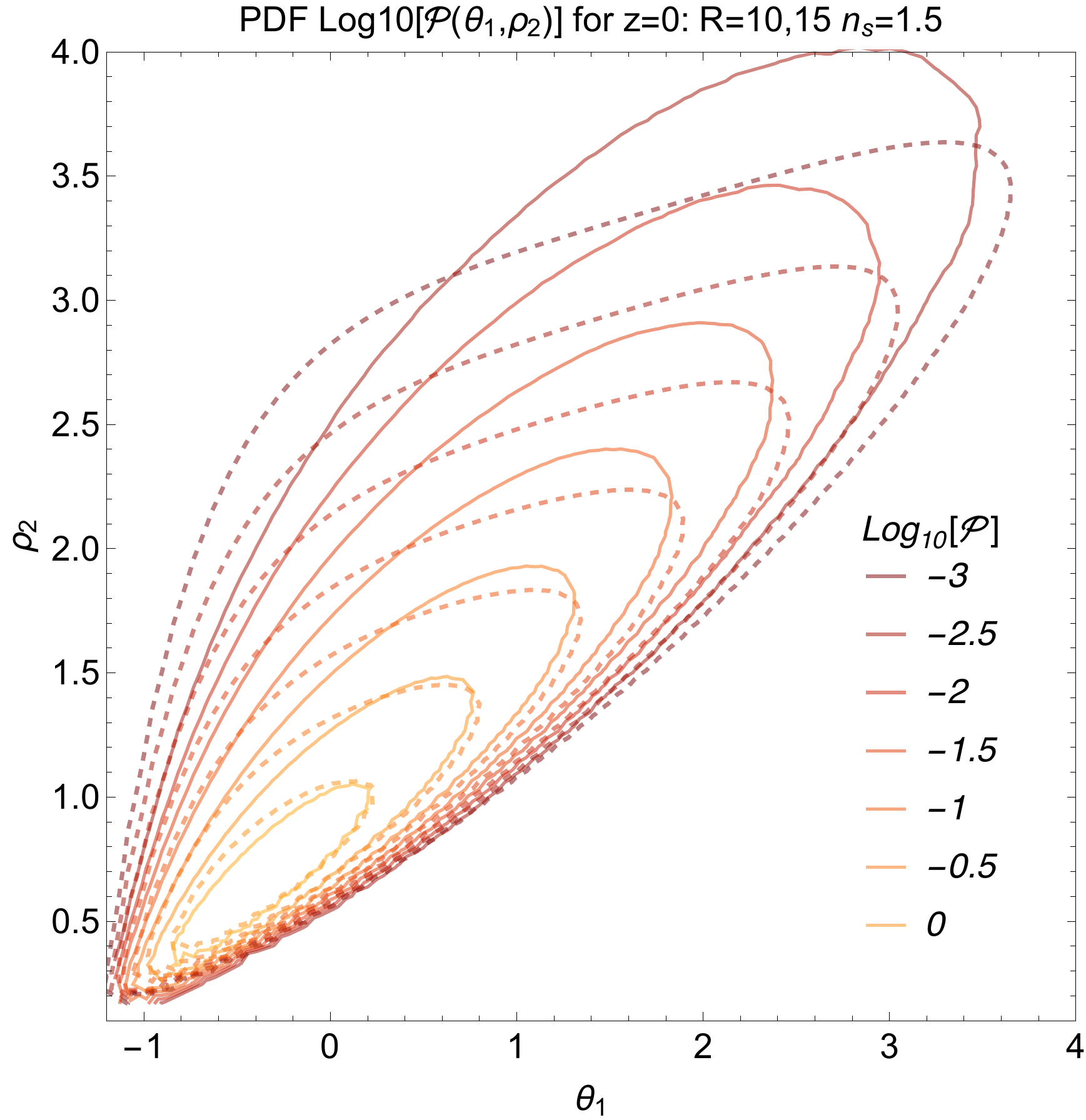}
\includegraphics[width=0.8\columnwidth]{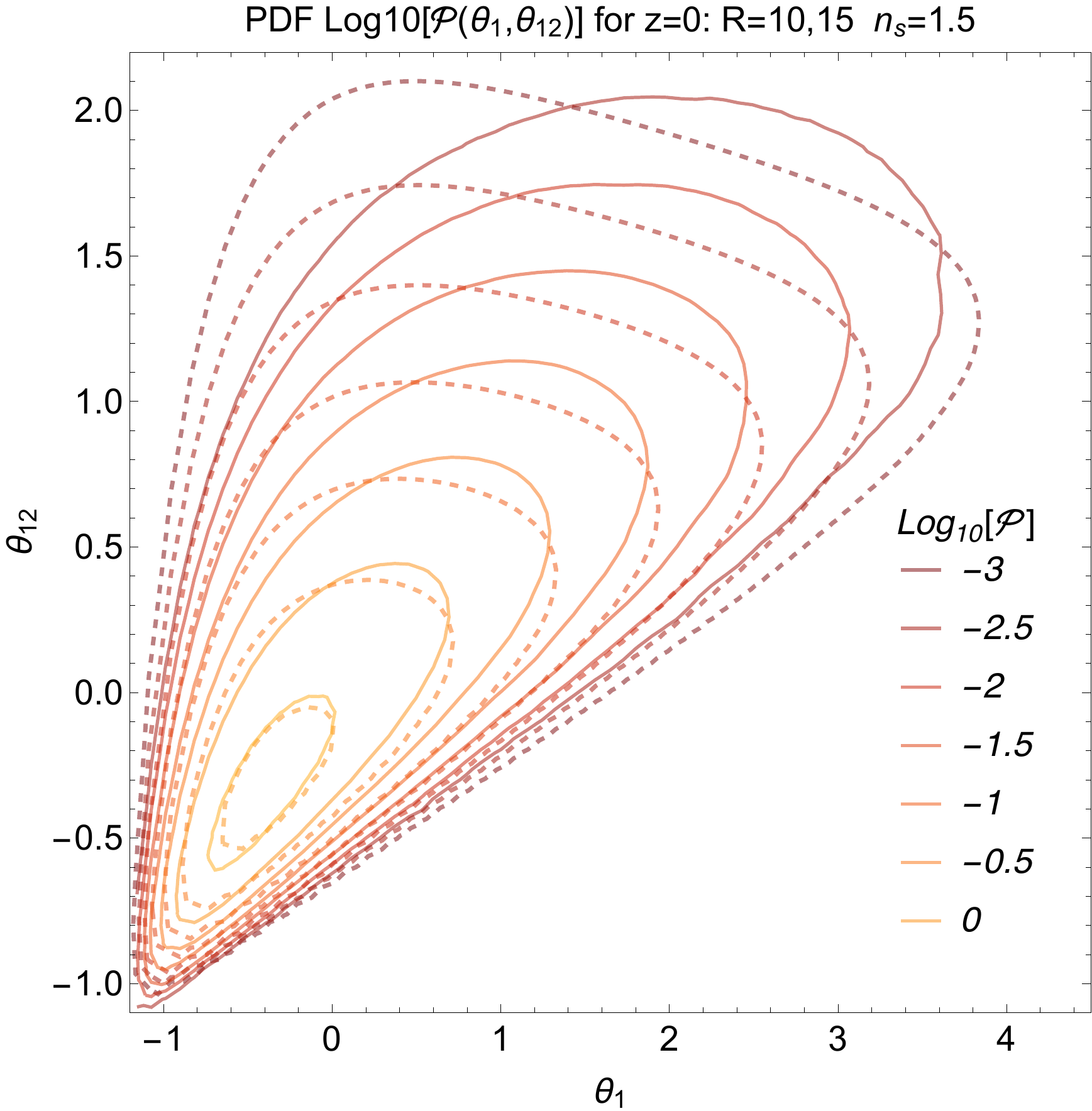}
\caption{Comparison of the joint PDF of the density $\rho_1$ and velocity divergence $\theta_2$ (upper panel), the velocity divergence $\theta_1$ and the density $\rho_2$ (middle panel) as well as two velocity divergences $\theta_{1}$ and $\theta_{12}=\theta_{1}-\theta_{2}$ (bottom panel) , 
for two different radii $R_1=10$ and $R_2=15$Mpc$/h$ measured in the $N$-body simulation (thin lines) with the saddle point approximations for $\mu_\rho=\log\rho(\theta)$ computed with a power-law initial power spectrum with index $n=-1.5$ (thick dotted lines).
{Contour lines scan decades of the probability distribution as labeled.}}
   \label{fig:PDF-2cell-theta-num-vs-theo}
\end{figure}

The joint PDFs of density and velocity divergence at two different scales obtained after the normalization described in Section~\ref{sec:norm} are shown in Fig.~\ref{fig:PDF-2cell-theta-num-vs-theo} for the case of mixed joint PDFs of density and velocity divergence and the pure joint PDF of the velocity divergence. For the mixed cases we find a very good agreement in the central regions of the PDF that degrades towards the tails of the distribution. The accuracy of the result is however rather good and similar to the accuracy found for the joint statistics of the two-scale density field in previous works \citep{Bernardeau15,Uhlemann16}. 
The residuals found, especially in the tails and for the pure joint PDF, most likely have physical origins such as correlations between scales beyond the spherical collapse model that require to incorporate tidal corrections. Note that, in order to rule out potential errors that stem from our assumptions (saddle-point approximation for the PDF and parametric expression for the variance), we performed two tests: i) tabulating numerically the variance of the true linear power spectrum and ii) computing numerically the inverse Laplace transform to get the PDF. In both cases, we did not find any evidence for a significant impact of resp. the running of the spectral index or the saddle-point approximation, at least within the error bars associated with those noisy methods. This suggests that, while the PDF of the velocity divergence was more robust against changes in the variance than the density, especially the joint PDF of two velocity divergences is indeed more sensitive to changes in the variance than the joint PDF of two densities that has been compared to simulations in \cite{Uhlemann16b}. Again, we demonstrate this property in Appendix~\ref{app:cumulants} based on joint cumulants obtained from perturbation theory that are shown in Figs.~\ref{fig:S112}~and~\ref{fig:T112}. Given the importance of tidal corrections that has been demonstrated in \cite{ScoccimarroFrieman96,FosalbaGaztanaga98,Ohta03} for the unsmoothed velocity divergence, it would be not surprising that those tidal effects can alter the joint cumulants, even if they tend to cancel for simple cumulants of density and velocity fields smoothed at one scale. For future work, it would be interesting to study the impact of tidal corrections on those joint cumulants for unsmoothed fields and investigate the potential subsequent erasure of those when smoothed fields are considered.

\subsubsection{The constrained PDF for velocity divergence given density}
\begin{figure}
\centering
\includegraphics[width=0.995\columnwidth]{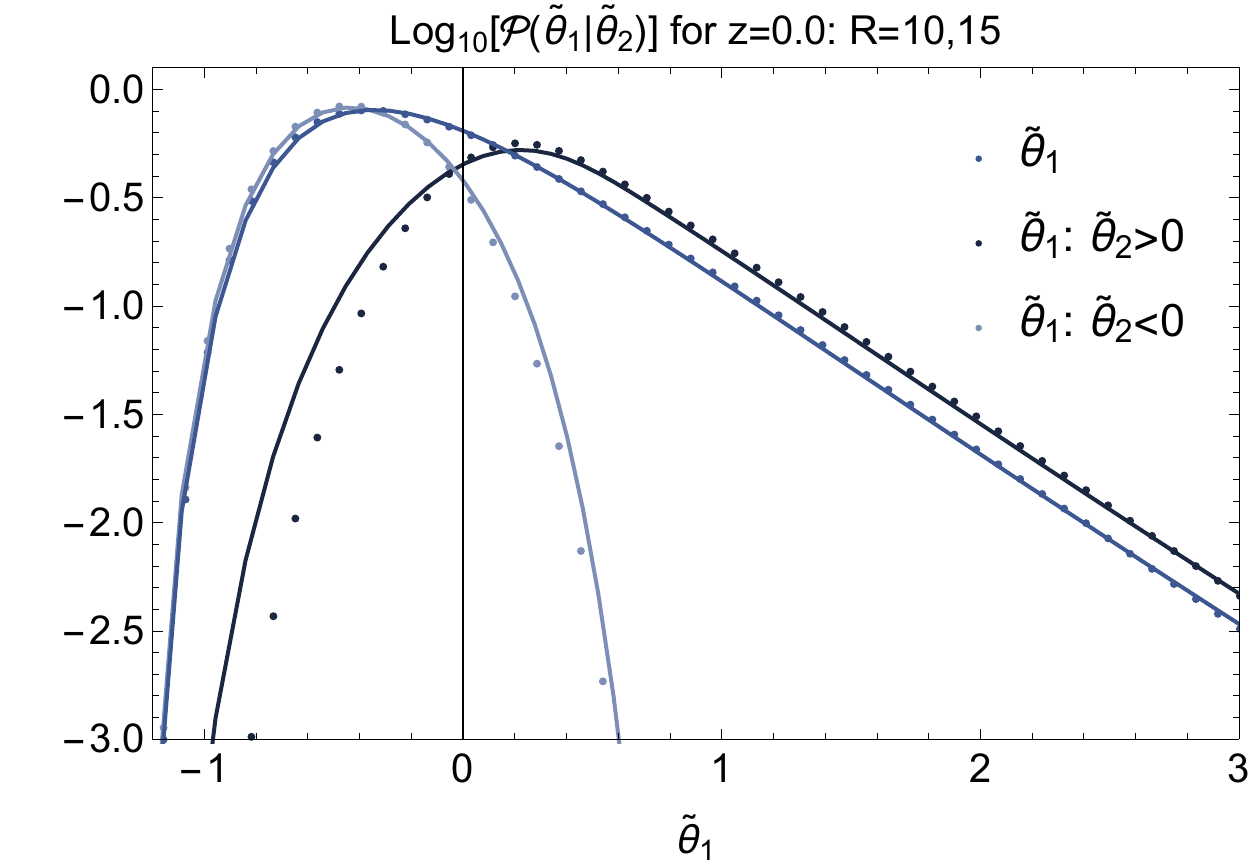}
\caption{Conditional PDF of the velocity divergence $\mP(\theta_1|\theta_2\gtrless 0)$ in an overdense (dark blue line) or underdense (light blue line) environment compared to the unconstrained case (blue line) derived from the two-cell saddle-point approximation in comparison to the measurements (data points). }
   \label{fig:condPDFtheta}
\end{figure}

The constrained PDF of the velocity divergence given an over- or underdense environment (or equivalently positive or negative velocity divergence) is obtained from the two-cell PDF by the following marginalization
\begin{align}
\label{eq:constrPDFveldif}
\mP(\tilde\theta_1|\tilde\theta_2\gtrless 0) &= \frac{\displaystyle \int_{-\nu}^\infty d\tilde\theta_2\ \Theta\left(\pm \tilde\theta_2\right) \mP(\tilde\theta_1,\tilde\theta_2)}{\displaystyle \int_{-\nu}^\infty d\tilde\theta_1 \int_{-\nu}^\infty d\tilde\theta_2 \ \Theta\left(\pm \tilde\theta_2\right) \mP(\tilde\theta_1,\tilde\theta_2) \ } \,,\\
\mP(\tilde\theta_1|\rho_2\gtrless 1) &= \frac{\displaystyle \int_{0}^\infty d\rho_2\ \Theta\left(\pm (\rho_2-1)\right) \mP(\tilde\theta_1,\rho_2)}{\displaystyle \int_{-\nu}^\infty d\tilde\theta_1 \int_{0}^\infty d\rho_2 \ \Theta\left(\pm (\rho_2-1)\right) \mP(\tilde\theta_1,\rho_2) \ } \,, \nonumber
\end{align}
where $\Theta$ is the Heavyside step function. The result that is obtained from the joint saddle point PDF by performing the integration in equation~\eqref{eq:constrPDFveldif} numerically, is shown in Fig.~\ref{fig:condPDFtheta}. The one-cell marginal $\mP(\theta_1)$ shown here agrees well with the direct prediction for one-cell which is shown in the upper right panel of Fig.~\ref{fig:PDF-1cell-rhotheta-num-vs-theo}.The agreement between the prediction of the conditional PDFs (solid lines) and the measurements in our simulation (dots) is very good in particular for similar velocity divergences. We observe however some departures for large slopes corresponding to large positive values of $\theta_{1}$ when $\theta_{2}$ is negative and vice versa. This feature was not observed for the density field and could come from the scatter of the velocity-density relation which is not accounted for in the spherical collapse model. 

\subsubsection{Conditional velocity profile given density}
\label{sec:conditional}
\begin{figure}
\includegraphics[width=1\columnwidth]{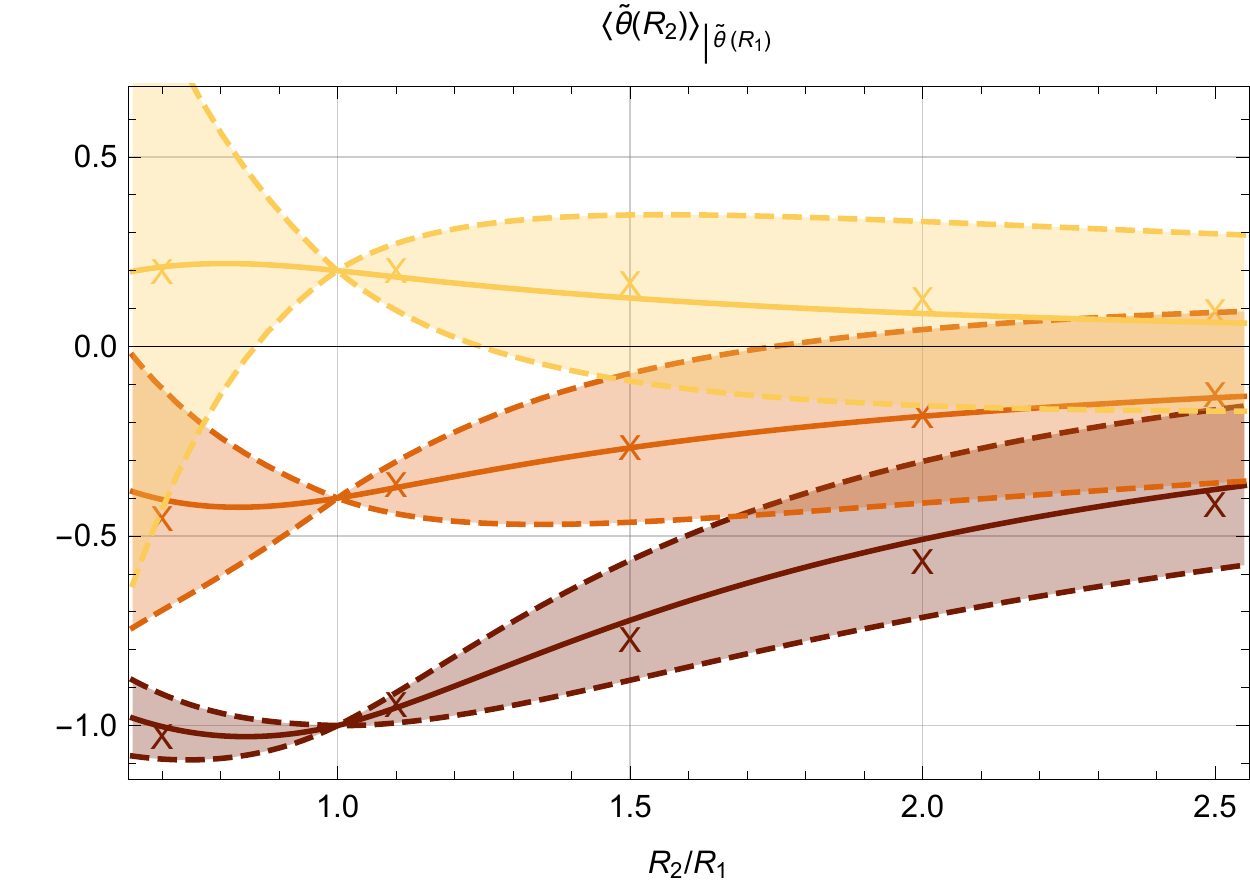}
\caption{Mean conditional profile $\langle \tilde\theta_2\rangle_{\tilde\theta_1} \simeq \langle \tilde\theta_2\rangle_{\rho_1}$ including the variance computed from the saddle-point approximation equation~\eqref{eq:condprofile} as a function of the ratio of radii $R_2/R_1$ where $R_1=10$ Mpc$/h$ for values of $\tilde\theta_1=\{-1.0, -0.4, 0.2 \}$ that corresponds to $\rho_1\approx \{0.2,0.6,1.2\}$ in comparison to the measured values for $R_2=\{7,15,20,25\}$ Mpc$/h$ (crosses). 
}   \label{fig:condprofiletheta}
\end{figure}

The saddle point approximation can be used to determine the conditional velocity profile, completely analogous to the density profile obtained in \cite{Bernardeau14}, by using the stationary condition for the decay-rate function
\begin{align}
\label{eq:condprofile}
0=\frac{\partial \Psi(\theta_1,\theta_2)}{\partial\theta_2} \Bigg|_{\theta_2=\bar\theta_2(\theta_1)}\,,
\end{align}
where the constrained variance is given by
\begin{align}
\label{eq:condprofilevariance}
\langle \theta_2^2 \rangle_{\theta_1} - \langle \theta_2 \rangle_{\theta_1}^2= \left[\frac{\partial^2 \Psi(\theta_1,\theta_2)}{\partial \theta_2^2} \right]^{-1} \Bigg|_{\theta_2=\bar\theta_2(\theta_1)} \,.
\end{align}
The result of this prediction is shown in Fig.~\ref{fig:condprofiletheta} in comparison to the measurements from the simulation which are in very good agreement with the theory. Note that, the conditional profile $\langle\theta_2\rangle_{\theta_1}$ corresponds to $\langle\theta_2\rangle_{\rho_1}$ if $\rho_1=\rho_{\rm SC}(\theta_1)$. However, the variance obtained from \eqref{eq:condprofilevariance} for the mixed case $\Psi(\rho_1,\theta_2)$ becomes problematic because contributions beyond tree level are important when similar radii are considered due to the scatter around the mean relation between the density and velocity divergence. 

\section{Application: Cosmic parameter estimation}
\label{sec:applications}
The fully analytical  probability density function for the mildly non-linear cosmic velocity field within
 spherical cells described in Section~\ref{sec:PDFconstruction} can be used jointly with the corresponding density prediction to build a simple but
 accurate maximum likelihood estimate for the redshift  evolution of the variance of both density (as first investigated by \cite{Codisetal2016}) and velocity divergence, which, can also be shown to have smaller relative errors than their sample variances. Note that our formalism in principle allows us to get a maximum likelihood estimate of all higher-order cumulants of the density and velocity divergence fields which would have again a smaller sample variance \citep[see e.g][for an estimate of the sample variance of the velocity kurtosis]{2000PASJ...52..749S}.

Indeed, a dark energy probe may directly attempt to estimate the so-called equation of state parameters $(w_a,w_{p})$ from the two PDF  while relying on
 the cosmic model for the growth rate  and the growth factor \citep{Glazebrook}, 
\begin{align}
D(z|w_{p},w_a)&=\frac{5\Omega_m H_0^2}{2} H(a)\int_0^a \frac{{\rm d} a'}{a'^3 H^3(a')}\,, \\
f(z|w_{p},w_a)&=\frac{{\rm d}\log D}{{\rm d}\log a}\,, \\
H^2(a)=H_0^2 &\left[ \frac{\Omega_m}{a^3}+ \Omega_\Lambda \exp\left(3 \int_0^z \frac{1+w(z')}{1+z'} {\rm d} z'  \right)
 \right], \, \label{eq:cosmo}
\end{align}
 with $\Omega_m$, $\Omega_\Lambda$ and $H_0$ resp. the dark matter and dark energy  densities and the Hubble constant at redshift $z=0$,  $a\equiv1/(1+z)$ the expansion factor, and with the equation of state $w(z)=w_p+ w_a /({1+z})$.

\subsection{A joint fiducial dark energy experiment}
\label{sec:DE_experiment}

Let us  extend the fiducial experiment presented in \cite{Codisetal2016} to account for the 
joint fit of both PDFs. Let us consider a survey with redshifts between 0.1 and 1 binned so that the comoving distance of one bin is $40$Mpc$/h$, and  let's draw regularly spheres of radius $R=10$Mpc$/h$ separated by $d=40$Mpc$/h$ (hence we ignore neighbouring spheres and assume that the spatial correlations are negligible as shown by \cite{CBP2016}).
For a 15,000 square degree survey in a flat sky approximation \footnote{Here we use the flat sky approximation and we draw spheres regularly on a grid for the sake of simplicity. To get more precise estimates, one could account for the curvature and check which configuration of spheres maximizes their total number while preventing them from being closer than $40$Mpc$/h$ from their neighbours.}, it yields 50 bins of redshift ($z_{i}$) with a number of spheres ranging from about $N_{1}=800$ (at $z_{1}=0.1$) to $N_{50}=45,000$ (at $z_{50}=1$) for a total of almost 900,000 supposingly independent spheres.
In this experiment, we assume that the model for the density and divergence PDFs are exact and that the variances (which are free parameters) are related to the growth rate  and the growth factor by linear theory.
At each redshift, we can reconstruct the variance of each PDF of $x$ (either $\rho$ or $\theta$) by measuring the full PDF
 \begin{equation}
\hat \sigma_{\rm ML}(z_{j})= \arg \max_{\sigma} \left\{\prod_{i=1 }^{N_{j}}   \mP\left(x_{i,j}|\sigma
 \right) \right\} \,.
 \end{equation} 
A typical  precision on $\sigma$ of a percent is found for each PDF.  As mentioned in  \cite{Codisetal2016}, the reconstruction is more accurate at higher redshift where the accessible volume and therefore the number of spheres is larger.

\begin{figure}
\includegraphics[width=\columnwidth]{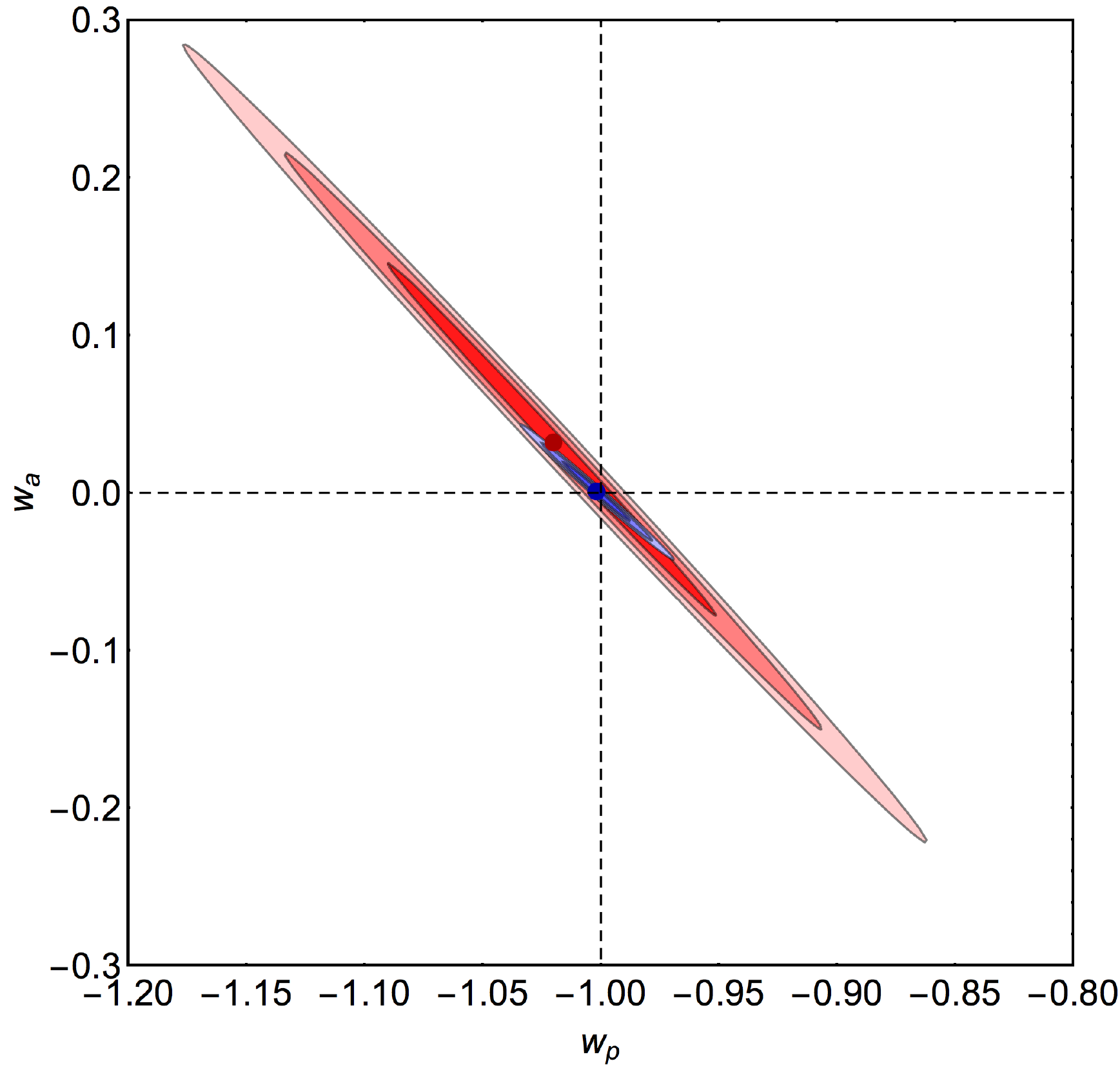}
   \caption{Contraints on the equation of state of dark energy in a Euclid-like survey containing about 900,000 spheres of comoving radius $10$Mpc$/h$ regularly drawn between redshift $0.1$ and $1$.  Contours at one, two and three sigmas are displayed with shaded areas from dark to light red when measurements of densities are done and dark to light blue for velocity divergence measurements. The dark red and blue dots correspond to the recovered most likely solution, the input being $(w_p,w_{a})=(-1,0)$. The resulting precision on $w_{p}$ and $1+w_{a}$ is respectively at about the
   $6\%$ and $10\%$ level for densities and $2\%$ and $2\%$ level for velocities.
   \label{fig:MLeos-Euclidlike}}
\end{figure}

In order to get constraints on the equation of state of dark energy, we compute the log-likelihood of the $\sim 900,000$ measured densities and velocities $\{\rho_{i,j},\theta_{i,j}\}_{{1\leq i\leq N_{j}},{1\leq j\leq 50}}$ given models for which $w_{p}$ and $w_{a}$ vary
\begin{align}
{\cal L}(\{x_{i,j}\}|w_{p},w_{a})&=\sum_{j=1}^{50}\!\sum_{i=1}^{N_{j}}\log {\cal P}_x(x_{i,j}|z_{j},w_{p},w_{a})\,, \nonumber 
\end{align}
where ${\cal P}_\rho(\rho|z,w_{p},w_{a})$ and  ${\cal P}_\theta(\theta|z,w_{p},w_{a})$  are resp. the theoretical density PDF for the density $\rho$ and 
the divergence $\theta$ at redshift $z$ for a cosmological model with dark energy e.o.s parametrised by $w_{p}$ and $w_{a}$.
Optimizing the probability of observing densities and divergences $\{\rho_{i,j},\theta_{i,j}\}_{{1\leq i\leq N_{j}},{1\leq j\leq 50}}$  at redshifts $\{z_{j}\}_{{1\leq j\leq 50}}$ with respect to  $(w_{p},w_a)$,  
yields a maximum likelihood estimate for the dark energy equation of state parameters
 \begin{equation}
 (\hat w_p,\hat w_a)= \arg \max_{w_p,w_a} \left\{{\cal L}(\{x_{i,j}\}|w_{p},w_{a}) \right\} \,. \label{eq:defDop2t}
 \end{equation}
The resulting $\alpha=1,2,3$ sigma contours are shown in Fig.~\ref{fig:MLeos-Euclidlike} and correspond to the models for which  ${\cal L}(\{x_{i,j}\}|w_{p},w_{a})=\max_{w_{p},w_{a}} {\cal L}(\{x_{i,j}\}|w_{p},w_{a}) +\log (1-{\rm Erf}(\alpha/\sqrt 2))$.
Modulo our assumptions, this maximum likelihood method  allows for constraints on $w_{p}$ at a few percent and $1+w_{a}$ at about a ten percent level when the PDF of the density is used,
making it a competitive tool for the analysis of future Euclid-like surveys as already pointed out by  \cite{Codisetal2016}.
Fig.~\ref{fig:MLeos-Euclidlike} compares results when only densities are used (in red) to the sole use of velocity data (blue).
As expected, for a same survey volume, the velocity divergence PDF allows for much tighter constraints on the equation of state of dark energy than the density PDF alone (by a factor of more than 10!). 

 In practice, surveys mapping the peculiar velocity field have a much smaller volume than those targeting the galaxy distribution but despite this drawback, they should still carry a competitive cosmological information.
It is expected that various uncertainties will degrade the accuracy of the proposed method (uncertainties on the model itself at low-redshift, galaxy biasing, redshift space distortions, observational biases, etc). 
Accounting for these effects requires further works, beyond the scope of this paper. 
Our main conclusion should still hold once all those effects are accounted for, namely that there is a  significant gain in using the full knowledge of both PDFs --  therefore relying on a maximum likelihood analysis -- when contrasted  to the direct measurements of cumulants (variance, skewness, etc).  Large-deviation theory will therefore allow us to get tighter cosmological constraints.

\subsection{$f(\Omega)$ and $\sigma_{8}$ from velocity PDF}
\begin{figure*}
\includegraphics[width=0.95\columnwidth]{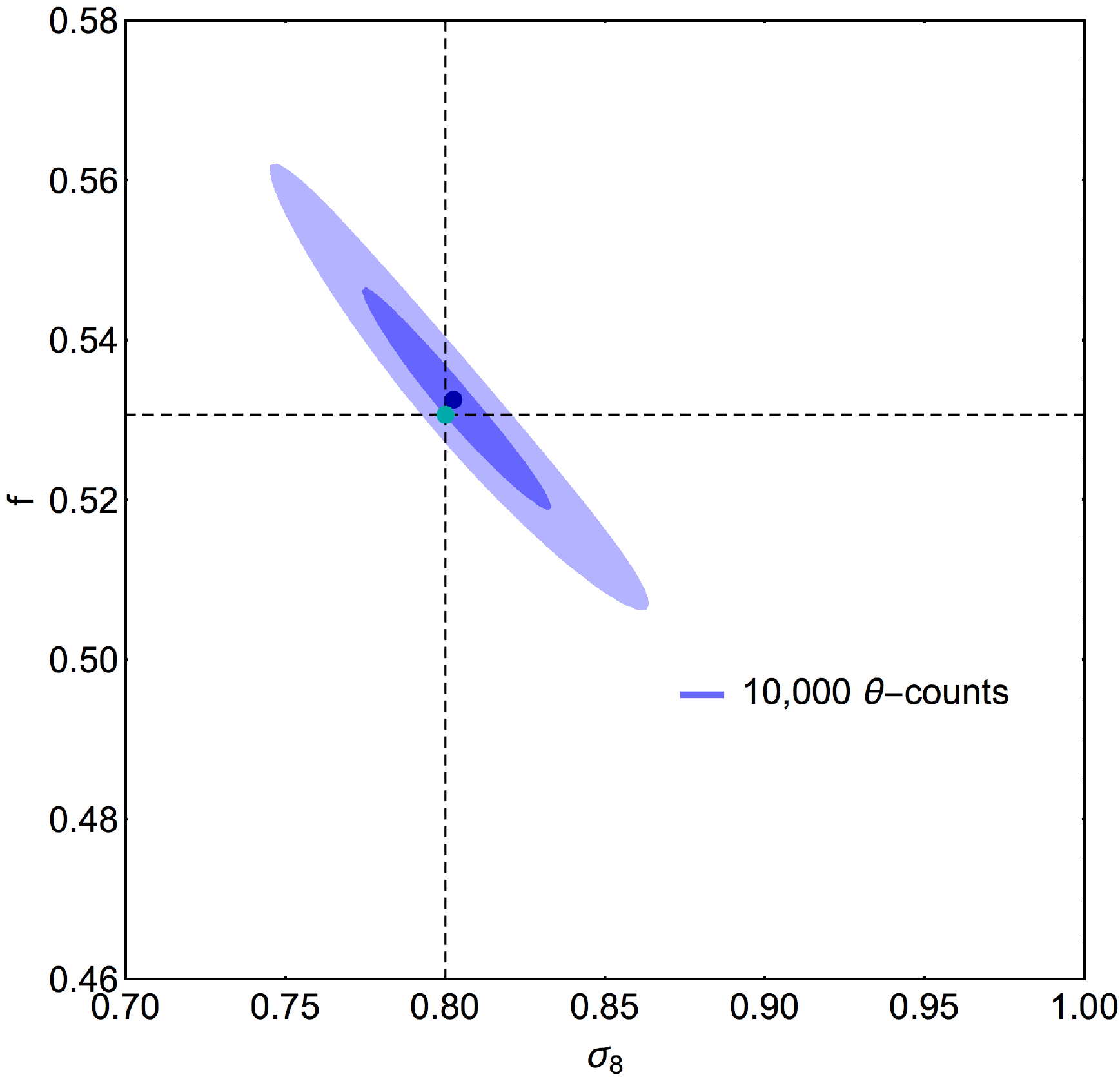}\hskip 0.5cm
\includegraphics[width=0.95\columnwidth]{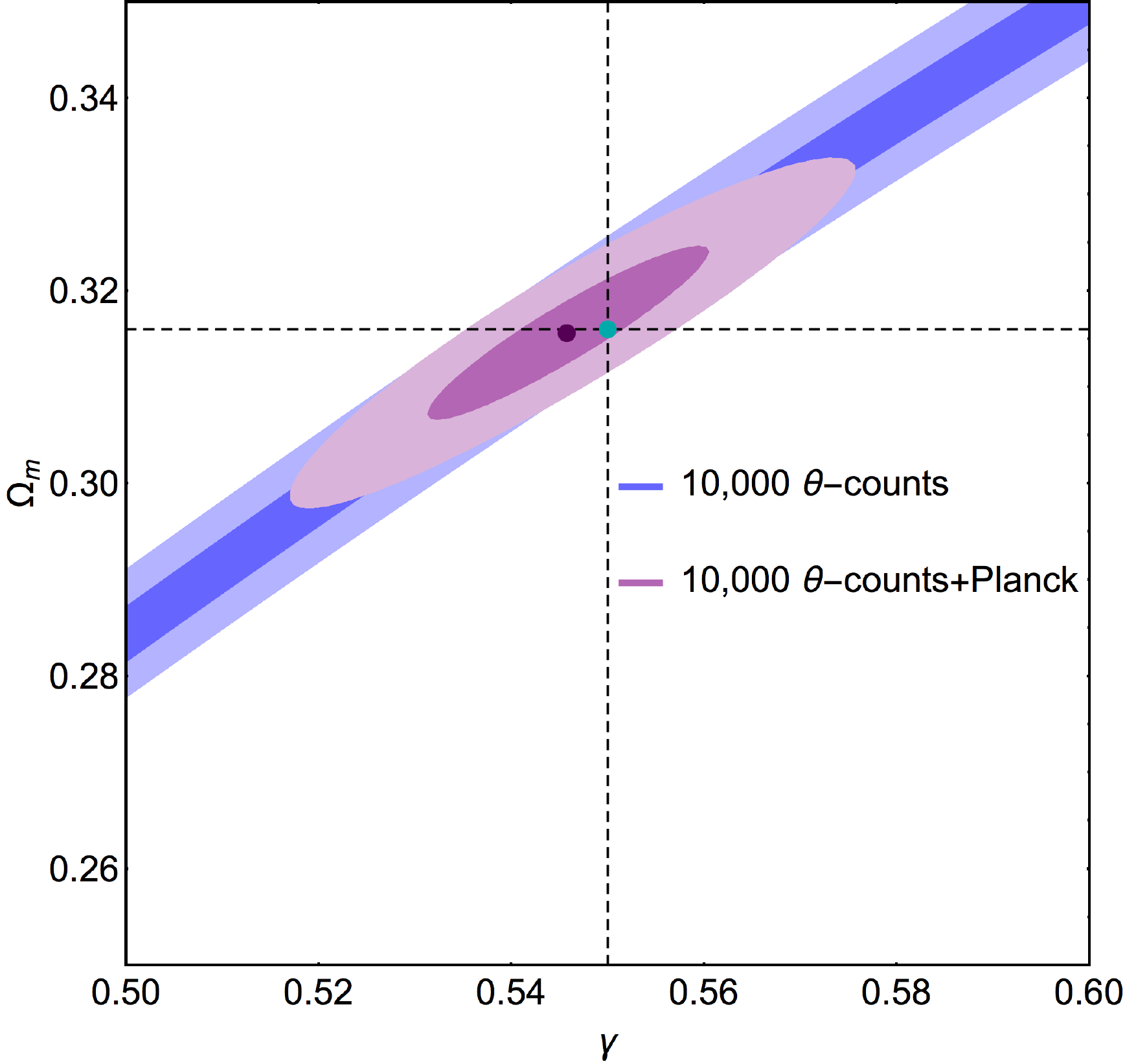}
   \caption{Left-hand panel: Constraints on $\sigma_{8}$ and $f(\Omega)$ obtained from velocity measurements in
10,000 spheres of 20Mpc$/h$ at $z=0$. Right-hand panel: same as left-hand panel when we add a prior from Planck on $\Omega$ to get constraints on $\gamma$. Input parameters are shown with the cyan dot while blue and purple dots show the maximum likelihood estimates. The one and two sigma contours are shown with the dark and light shaded areas. 
   }
   \label{fig:MLeos-exp1}
\end{figure*}
Alternatively and to avoid uncertainties due to galaxy biasing, we propose another experiment in which we use velocity data alone. Our model for the PDF of the velocity divergence then depends on the spherical collapse dynamics that is assumed to be fixed here, $f(\Omega)$ -- which is now allowed to vary -- and the linear power spectrum whose amplitude will be denoted $\sigma_{8}$ as usual and redshift evolution is governed by dark energy following Equation~(\ref{eq:cosmo}). Measuring the velocity divergence PDF in the local Universe then provides low-redshift constraints on $\sigma_{8}$ and $f(\Omega)$ directly (here we will assume that dark energy is a cosmological constant). To illustrate this point, we measure the mean velocity divergence in 10,000 spheres of 20Mpc$/h$ at $z=0$. The corresponding constraints on $\sigma_{8}$ and $f(\Omega)$ are shown on the left-hand panel of Fig.~\ref{fig:MLeos-exp1} which demonstrates that using the full statistics of the velocity divergence field (beyond the sole skewness as proposed e.g by \cite{1995MNRAS.274...20B}) can allow us to get interesting constraints. Adding a Planck prior on $\Omega$ in addition allows us to get an estimate of $\gamma$ and therefore could be used to constraint modified gravity models.

In this work, we have assumed Gaussian initial conditions but it should be straightforward to include non-Gaussian initial conditions. In this case, the velocity divergence PDF could allow us to measure precisely the departure from Gaussian initial conditions, improving upon previous studies which focused on a subset of cumulants like the skewness \citep{1997MNRAS.286..223P} by gathering information from the full PDF (i.e all cumulants of the field).

\subsection{Velocity divergence and the extraction of bias}
If we assume that the count-in-cell statistics of both the density and the velocity divergence are well-described by large deviation statistics together with the spherical collapse mapping, we obtain an effective mapping between the two PDFs $\mP_{\tilde\theta_m} \stackrel{\rm SC}{\longleftrightarrow} \mP_{\rho_m} $. In Figs. \ref{fig:cond-PDF-rho-theta-sameR} and \ref{fig:PDF-1cell-rho-from-theta} we already demonstrated that the density PDF can be reconstructed from the PDF of the velocity divergence using the spherical collapse mapping. For the simulation under consideration, at redshift $z=0$ and radius $R=10$ Mpc/$h$, the accuracy of this reconstruction is within 10\% for densities between 0.3 and 3. Having established this relation between the statistics of the density and velocity field for dark matter now gives us a handle on comparing the statistics of the density of dark matter and halos. By assuming that velocities between matter and halos are unbiased $\mP_{\tilde\theta_h} \stackrel{\tilde\theta_h=\tilde\theta_m}{=} \mP_{\tilde\theta_m}$ we can obtain an effective measurement of the dark matter density PDF by measuring halo velocities. On the other hand we can directly measure the halo density PDF which finally allows us to determine the bias between matter and halos. The idea can be summarised as follows
\begin{equation}
\mP_{\tilde\theta_h} \stackrel{\tilde\theta_h=\tilde\theta_m}{=} \mP_{\tilde\theta_m} \stackrel{\mathrm {\rm SC}}{\longleftrightarrow} \mP_{\rho_m}  \stackrel{\delta_h=\sum_i b_i\delta_m^i}{\boldsymbol{\longleftrightarrow}} \mP_{\rho_h} \,.
\end{equation}
Since the reconstruction of the density from the velocity dispersion is accurate around the peak of the distribution, which is significantly transformed due to the effect of bias (Feix et al in prep.), we expect that measuring the PDF of the velocity divergence along with the PDF of the biased dark matter field will allow to extract properties of bias.

\subsection{Measurements of peculiar velocities from real surveys}
In galaxy redshift surveys, the observables that are related to the radial peculiar velocity (along the line of sight) are the redshifts and radial distances of galaxies $cz=H_0r-v_r$. While the redshift can be obtained accurately from spectroscopy, extracting a peculiar velocity field also requires using  empirical distance indicators such as the Tully-Fisher relation for spiral galaxies \citep[][relating the luminosity to the angular velocity $L\propto v^4$]{TullyFisher77}, the Faber-Jackson relation \citep[][relating the luminosity to the central stellar velocity dispersion $L\propto \sigma^4$]{FaberJackson76} for elliptical galaxies  and extended versions thereof such as the Fundamental Plane \citep{DjorgovskiDavis87} or luminosity-distance relation for Type Ia supernovae \citep{Tonry03}. Note that those traditional ways of estimating distances have a distant-dependent uncertainty and it gets increasingly difficult to estimate velocities for deep samples and large volumes. One can avoid this problem by using the CMB temperature distortion that is induced by the kinetic Sunyaev-Zeldovich effect (the scattering of CMB photons by the ionized gas in intervening structures which Doppler shifts wavelengths depending on the line-of-sight peculiar velocity) which allows to extract peculiar velocities for galaxy clusters \citep{Hand12}, but due to a weakness in signal and the contamination by primary CMB fluctuations this method does not seem to be promising for detecting peculiar velocities of individual galaxies.\\
All those methods have in common that they only give the line-of-sight component of the peculiar velocity field. However, the three-dimensional velocity field can be reconstructed using methods like POTENT \citep{BertschingerDekel89}, Least Action \citep{Peebles89} or Monge-Amp\`ere-Kantorovich (MAK) \citep{Brenier03}. We believe that the underlying assumption, that the velocity can be well approximated as a gradient field, is not an obstruction in our case given that our statistical estimator is for average velocity divergences in spheres such that the radial component of the peculiar velocity suffices and nonlinear effects of vorticity on very small scales are irrelevant \cite{PichonBernardeau99}.

\section{Conclusions}
\label{sec:conclusion}

The general formalism to describe the cosmic evolution of the joint statistics of the density and the velocity divergence fields in multiple concentric spheres was presented, and validated against state-of-the-art numerical estimators for the velocity divergence. 

In particular, we obtained fully analytical predictions for the PDF of the velocity divergence in a sphere that are accurate at the 2\% percent level for a wide range of velocity divergences that extend beyond the range in which the density PDF has a comparable accuracy. It  was made possible through advances in the analytical theory for predicting PDFs as well as an improvement in simulations allowing to accurately resolve velocity fields. We also obtained the first theoretical prediction for the joint PDFs of densities and velocity divergences in spheres of two different radii and compared it to the simulation finding qualitatively good agreement. We used this result to compute the conditional PDF of finding a velocity divergence given a surrounding over or underdense region and the corresponding conditional velocity profile that is of interest for studies of both halos and voids.

To highlight the practical implications of our results, we built a  maximum likelihood estimator for the joint redshift evolution of the variance of the density and the velocity divergence that allows to constrain dark energy considerably better than using density fields alone. Finally, we provided an outlook of how the relationship between the density and velocity divergence in spheres and hence their PDFs might be used to estimate bias. For the future, it would be interesting to investigate the feasibility of a removal of redshift space distortions using the density-velocity relationship based on upcoming results for the PDF of the biased density fields {(Uhlemann et al in prep.)}. It could also be of interest to also revisit the two-point large separation limit for the bias function presented in \cite{CBP2016,Uhlemann16b}
 to predict the pairwise peculiar velocities function in the mildly non linear regime. 
 
\vskip 0.5cm
{\sl Acknowledgements:}  This work is partially supported by the grants ANR-12-BS05-0002 and  ANR-13-BS05-0010 of the French {\sl Agence Nationale de la Recherche}.  CU is supported by the Delta-ITP consortium, a program of the Netherlands organization for scientific research (NWO)  funded by the Dutch Ministry of Education, Culture and Science (OCW). OH acknowledges funding from the European Research Council (ERC) under the European Union's Horizon 2020 research and innovation programme (grant agreement No. 679145, project `COSMO-SIMS'). We warmly thank Martin Feix, Juhan Kim and Dmitri Pogosyan for discussions and the anonymous referee for useful suggestions.

\bibliography{LSStructure}


\appendix
\section{LSSFast Package}
\label{sec:code}

The density PDF of $\tilde\theta$ for power-law and arbitrary power spectra are made available in the updated  \href{http://cita.utoronto.ca/~codis/LSSFast.html}{LSSFast}  package distributed online.

\section{Construction of the PDFs}
\label{app:PDF}

\subsection{The large-deviation principle}
Using large-deviations theory, PDFs can be constructed following three steps.

\paragraph*{Step 1: Spherical collapse map to the final rate function}
The final decay-rate function can be obtained from the initial one using the so-called contraction principle saying that any large deviation follows the least unlikely of all unlikely transformations between the initial and final state. When considering statistics of spherically-averaged quantities the leading order contribution to the time evolution can be identified  as the dynamics of spherical collapse. This one-to-one mapping between the initial density contrast $\tau_k$ in a sphere of radius $r_k$ and the final density and  velocity divergence $(\rho_k,\theta_k)=(\rho_{\rm SC},\theta_{\rm SC})(\tau_k)$ within radius $R_k$ (with $r_k=\rho_k^{1/3} R_k$ due to mass conservation)  allows to determine the final decay-rate function as
\begin{equation}
\Psi_{f}(\Phi\!=\!\{(\rho,\theta)_k\})\!=\!\Psi_{i}\left(\!\{\tau_k\!:\!(\rho,\theta)_k\!=\!(\rho_{\rm SC},\theta_{\rm SC})(\tau_k)\}\!\right).
\end{equation}

\paragraph*{Step 2: Legendre transform to the cumulant generator}
The cumulant generating function (CGF) $\varphi$ is obtained as a Legendre transform of the decay-rate function $\Psi(\Phi)$ according to Varadhan's theorem:
\begin{equation}
\varphi(\Lambda)=\ltrans\Lambda \cdot \Phi-\Psi(\Phi)\,,
\end{equation}
where
\begin{equation}
\Lambda=\{\Lambda_k\} = \{ \nabla_{\Phi_{k}}\Psi(\Phi) \} \,.
\end{equation}

\paragraph*{Step 3: Inverse Laplace transform to the PDF}
Once the cumulant generating function $\varphi(\Lambda)$ is known, the PDF $\mP( \Phi)$ can be computed 
from the usual inverse Laplace transform
\begin{equation}
\label{eq:InvLapTransnD}
\mP(\Phi)=
\int_{-\imath\infty}^{+\imath\infty}\frac{\dd\Lambda}{(2\pi \imath)^{2N}}
\exp(-\ltrans\Lambda \cdot \Phi+\varphi(\Lambda)) \,,
\end{equation}
which requires an integration in the complex plane.

\subsection{Saddle point approximation}
 
 In general, the numerical integration in the complex plane of equation~(\ref{eq:InvLapTransnD}) can be challenging because of numerical oscillations and singularities. Indeed, the cumulant generating function often presents poles and branch cuts as underlined e.g by \cite{Bernardeau14} due the fact that the first derivative of the rate function has a maximum beyond which $\Lambda$ is not defined. An alternative to the brute force numerical integrations  performed in \cite{Bernardeau14,Bernardeau15} is to rely on a saddle point approximation when the singularity is sufficient far away from the values of interest. Depending on the variable used in the saddle-point approximation, the domain of validity of this analytical approximation vary.
 
Let us now describe how the complex integral can be obtained from a saddle point approximation that, using a well-chosen variable, provides analytical results as good as the numerical integration. 
Let us first call the appropriate set of variables for the general $N$ cell case $\{\mu_k\}$, we will later give the concrete variables that are suited to obtain good predictions for the joint density and velocity statistics. 
A saddle-point approximation of equation~\eqref{eq:InvLapTransnD} allows for a direct shortcut from step 1 to step 3. Indeed,
applying the saddle point approximation gives the following PDF 
\begin{subequations}
 \label{eq:saddlePDFN-cell}
 \begin{align}
 \label{eq:saddlePDFlogN-cell}
\mP_{\mu,\{R_k\}}(\{\mu_k\})&= \sqrt{\det\left[\frac{\partial^{2}\Psi_{\{R_k\}}}{\partial \mu_{i}\partial \mu_{j}}\right]} \frac{ \exp\left[-\Psi_{\{R_k\}}\right]}{ (2 \pi)^{N/2}} 
\,,
 \end{align}
 which does not require the explicit calculation of the Legendre transform described in step 2. From equation~(\ref{eq:saddlePDFlogN-cell}), it is clear that
the variables $\{\mu_k\}$ have to be chosen to ensure the convexity of the decay-rate function (${\partial^{2}\Psi_{\{R_k\}}}/{\partial \mu_{i}\partial \mu_{j}}>0$) which in practice boils down to pushing the singularity back to infinity. Equation~(\ref{eq:saddlePDFlogN-cell}) can then  be translated in the PDF of the density or velocity field $(\rho,\theta)=(\Phi^1,\Phi^2)$ via a simple change of variables
\begin{align}
\mP_{\{R_k\}}(\{\Phi_k^a\}) &=\mP_{\mu,\{R_k\}}[\{\mu_k(\{\Phi^a_i\})\}] \left|\det\left[\frac{\partial\mu_{i}}{\partial \Phi^a_{j}}\right]\right|  \,,
\label{eq:Prhofrommu}
\end{align}
where the Hessian of the decay rate function $\Psi_{\{R_k\}}$ after a change of variables $\{\Phi^a_k\}\rightarrow\{\mu_k\}$ is given by
\begin{align}
\frac{\partial^{2}\Psi_{\{R_k\}}}{\partial \mu_{i}\partial \mu_{j}}=\frac{\partial\Phi^a_k}{\partial\mu_i}\cdot \frac{\partial^{2}\Psi_{\{R_k\}}}{\partial \Phi^a_{k}\partial \Phi^a_{l}}\cdot \frac{\partial \Phi^a_l}{\partial\mu_j} + \frac{\partial^2\Phi^a_k}{\partial\mu_i\partial\mu_j}\cdot \frac{\partial\Psi_{\{R_k\}}}{\partial\Phi^a_k} \,.\notag
\end{align}
\end{subequations}

\subsection{Optimizing the reach of the saddle point approximation}
After introducing the general formula for the saddle point approximtion applied to a set of variables $\{\mu_k\}$, we now  present  exemplary results for different choices of variables for the one-cell case. This demonstrates the optimality of the log-density and log-mass mapping we use for the predictions and the comparison to the simulation in Section~\ref{sec:validation}.

The reach of the saddle point approximation applied to the density $\rho$ and the log-density $\mu_\rho=\log\rho$ has been extensively compared in \cite{Uhlemann16}, finding that the reach of the approximation for the log-density covers the whole range of densities that are of interest, while the application to the density leads to critical points that prevent its practical use. Inspired from this finding, we will generalise it to obtain the PDF of the velocity divergence $\theta$.

\label{sec:one_cell}
\begin{figure}
\centering
\includegraphics[width=1\columnwidth]{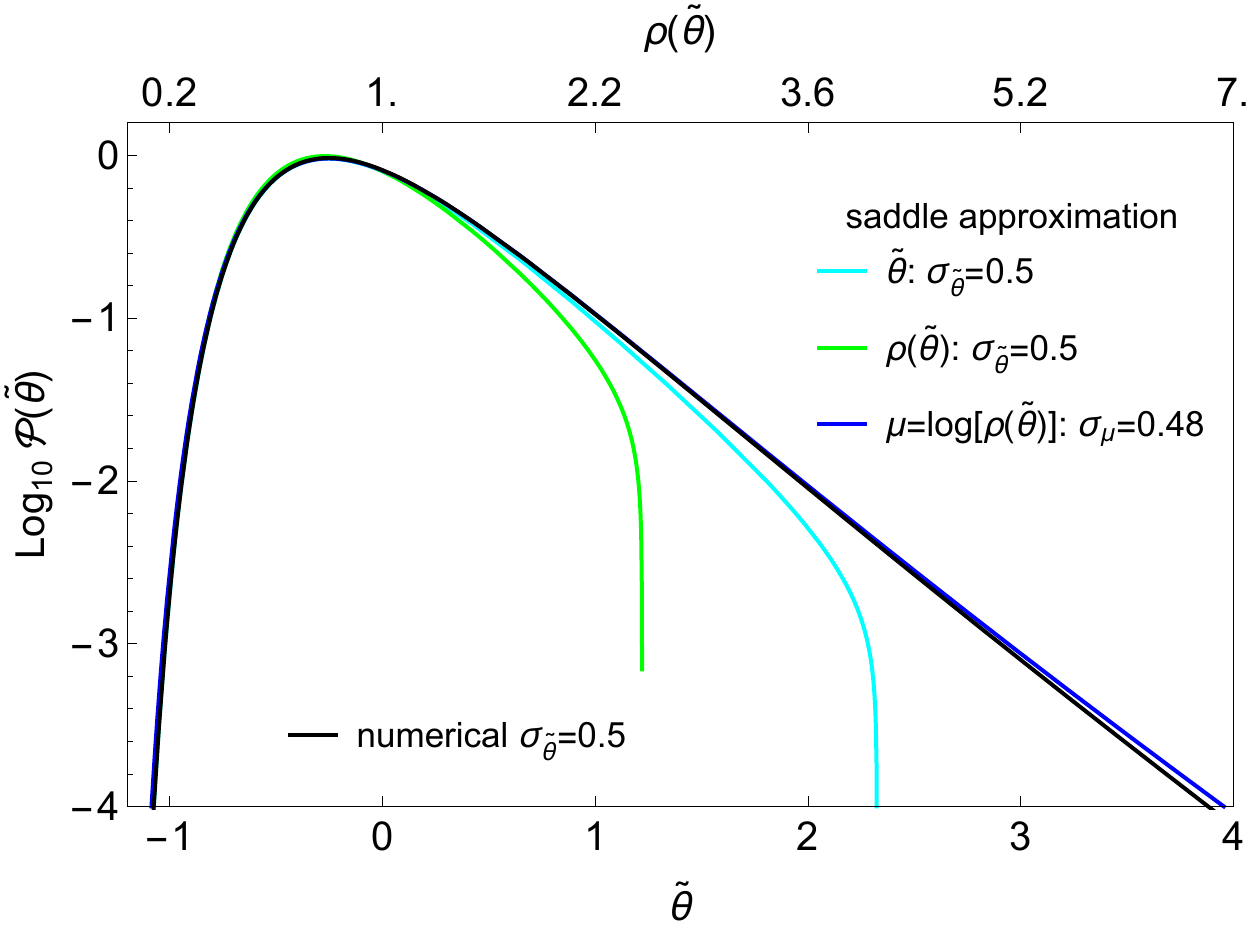}
   \caption{
Comparison of the reach of the saddle point approximation for the PDF of the velocity divergence $\tilde\theta$. The normalised saddle point PDF from a logarithmic transform according to equation~\eqref{eq:logrhofromtheta} {\it (blue line)} gives the same result as the numerical integration in the complex plane for $\tilde\theta$ {\it (black line)} in the whole range under consideration while the saddle point approximations for the density {\it (green line)} and velocity divergence {\it (cyan line)} fail earlier.
   }
   \label{fig:reachsaddlePDF}
\end{figure} 

In analogy to the PDF of the density, the saddle point approximation can be also applied with the velocity divergence as variable such that we have $\mu=\theta$ and hence
\begin{align}
\label{eq:saddlePDFtheta}
\mP_{R}(\tilde\theta) = \sqrt{\Psi''_R[\tilde\theta]/(2\pi)} \exp\left(-\Psi_R[\tilde\theta]\right)\,.
\end{align}
As for the saddle point approximation of the density PDF considered in \cite{Bernardeau14,Bernardeau15} this approximation fails as soon as the decay-rate function $\Psi_R(\tilde\theta)$ becomes non-convex. Although the saddle point approximation for the velocity divergence has a larger range of validity than the one for the density (see the cyan and green lines in Fig.~\ref{fig:reachsaddlePDF}), eventually criticality prevents its use for the whole range of values of interest.

However, as has been shown for the density in \cite{Uhlemann16} the criticality of the decay-rate function can be circumvented by using a logarithmic mapping hence allowing for fully analytical predictions of the PDF based on a saddle point approximation. Indeed, the PDF of the velocity divergence can be obtained by applying a saddle point approximation to the log-density which is in turn related to the velocity divergence by a one-to-one mapping
\begin{align}
\label{eq:logrhofromtheta}
\mu_{\rho(\tilde\theta)}=\log\left(\rho_{\rm SC}(\tilde\theta)\right) \,.
\end{align}
Note that this procedure is completely equivalent to using a logarithmic mapping for $\nu+\tilde \theta$. 

In Fig.~\ref{fig:reachsaddlePDF} we compare the range of validity for the saddle point approximation applied to different variables and compare them to the direct result for the PDF obtained from a numerical integration of equation~\eqref{eq:InvLapTransnD}. We find that while the saddle point approximations applied to the density or velocity divergence fail early, the log-density is valid in the whole range of interest and in very good agreement with the numerical integration. 

\section{Dispersion  and density relation}
\label{app:dispersion}

\begin{figure}
\centering
\includegraphics[width=1.0\columnwidth]{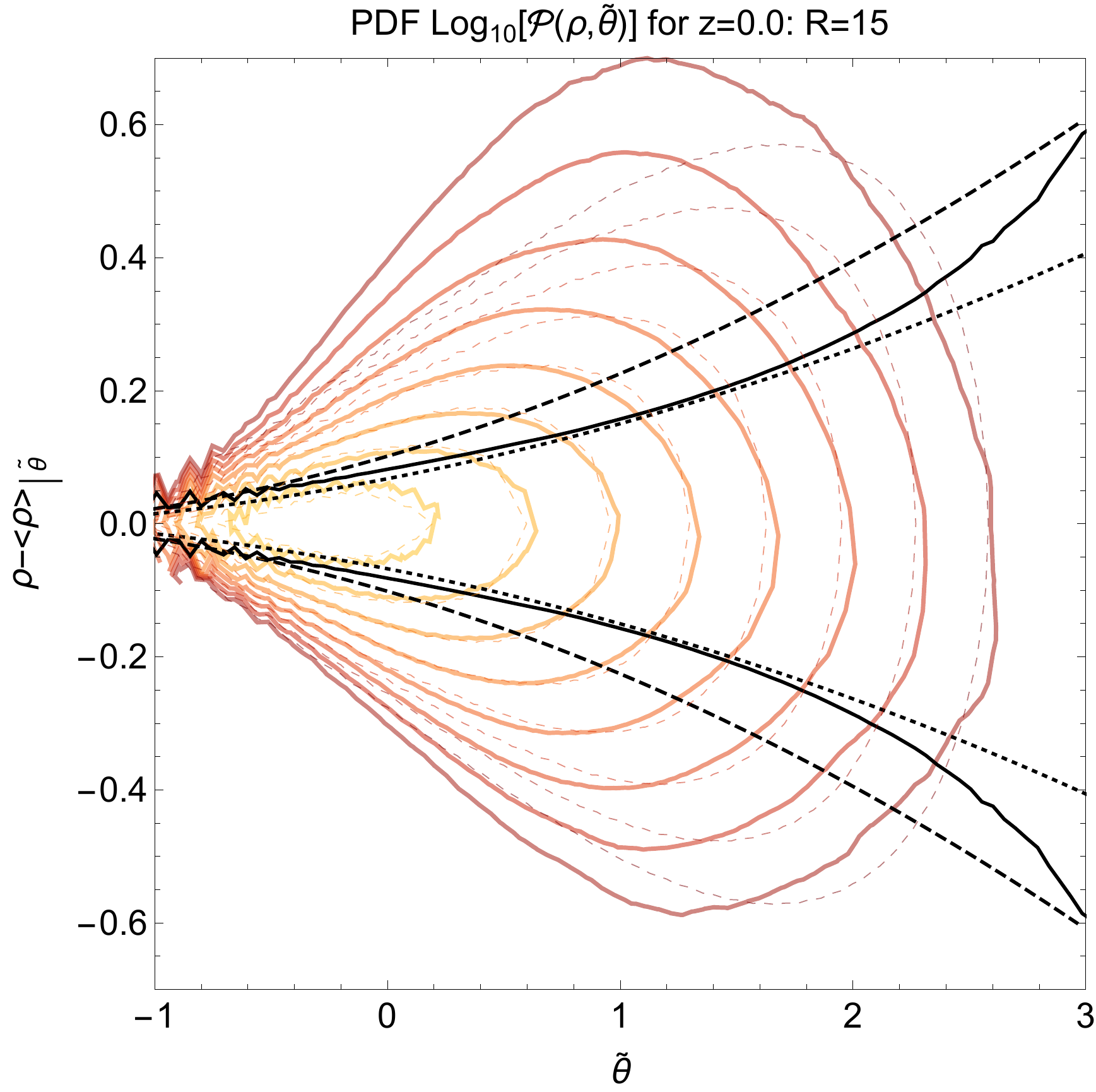}\\
   \caption{Joint PDF of the velocity divergence (bold coloured lines) and the scatter around the measured mean $\langle\rho\rangle_{|\tilde\theta}$ together with the typical scatter from the simulation (black line) and the fit~\eqref{eq:scatterrhofit} from \citep{BernardeauChodorowski99} (black dashed line) together with our improved fit (black dotted line) and the corresponding Gaussian model for the joint PDF (thin coloured lines) from equation~\eqref{eq:jointPDFrhothetasameR}. Note that the region $\tilde\theta\sim -1$ is completely dominated by a binning effect as the bin we use is approximately $\Delta \tilde\theta=0.08$.}
   \label{fig:joint-PDF-theta-scatterrho-sameR}
\end{figure}

 The scatter around the mean relation between density and velocity divergence can be obtained from conditional averages of 
\begin{align}
\sigma_{\rho|\tilde\theta}^2 &= \langle (\rho- \langle\rho\rangle_{|\tilde\theta})^2\rangle_{|\tilde\theta}\,,
\end{align}
They can also be expanded in perturbation theory up to third order
\begin{align}
\sigma_{\rho|\tilde\theta}^2 &= b_0 \sigma_{\tilde\theta}^4 +  b_2 \sigma_{\tilde\theta}^2\tilde\theta ^2 + b_1 \sigma_{\tilde\theta}^4\tilde\theta \,,
\end{align}
which has been computed at leading order $\mathcal O (\sigma^4)$ in \cite{Chodorowski98}. Since $b_2$ is identically zero for a top-hat filter,  \cite{BernardeauChodorowski99} argued that the next-to-leading order $\mathcal O (\sigma^5)$ can be of importance and included it in a phenomenological fit against numerical simulations finding
\begin{align}
\label{eq:scatterrhofit}
\sigma_{\rho|\tilde\theta} &= b_0 \left(1+\tilde\theta+\frac{2}{9} \tilde\theta^2\right)\sigma_{\tilde\theta}^2\,,
\end{align}
with $b_0=0.45$.
The scatter around the mean found numerically is shown in Fig.~\ref{fig:joint-PDF-theta-scatterrho-sameR} (black solid line) together with the fit from equation~\eqref{eq:scatterrhofit} using the original coefficient $b_0=0.45$ (dashed line) and our modified fit that uses $b_0=0.3$ (dotted line) and has been checked to give good results for smoothing radii $R=10, 11$ and $15$ Mpc$/h$. 
Interestingly, $0.3^2=0.09$ is much closer to the value predicted by perturbation theory that, depending on the spectral index, is between $b_0\simeq 0.05-0.09$ for smoothing radii of interest here.  If one assumes that the mean and the scatter completely characterise the joint distribution, one can make the following ansatz for the joint PDF of density and velocity divergence at the same scale R
\begin{align}
\label{eq:jointPDFrhothetasameR}
P_R(\rho,\tilde\theta) = \frac{1}{\sqrt{2\pi\sigma_{\rho|\tilde\theta}^2}} \exp\left[-\frac{(\rho-\langle\rho\rangle_{|\tilde\theta})^2}{2\sigma_{\rho|\tilde\theta}^2}\right] P_R(\tilde\theta)\,,
\end{align}
where the appropriate PT-inspired fit has to be inserted for $\sigma_{\rho|\tilde\theta}$ according to equation~\eqref{eq:scatterrhofit}. We show the result in comparison to the simulation measurements in Fig.~\ref{fig:joint-PDF-theta-scatterrho-sameR}.

\section{Cumulants of the  divergence }
\label{app:cumulants}

At tree-order (i.e vanishing variance), the reduced cumulants $T_n=\langle\tilde\theta^n\rangle_c/\sigma_{\tilde\theta}^{2(n-1)}$ of the velocity divergence field smoothed with a top-hat filter can all be obtained by applying a large-deviation principle to get the scaled cumulant generating function. In particular, for an Einstein-de Sitter Universe, its skewness and kurtosis read \citep{Bernardeau94smoothing}
\begin{eqnarray}
T_{3}&=&\frac {26}{7}+\gamma_{1}(R)\label{eq:T3},\\
T_{4}&=&\frac{12088}{441} + \frac{338}{21}\gamma_{1}(R)+ \frac{7}{3} \gamma_{1}(R)^2 + 
\frac 2 3 \gamma_{2}(R),\label{eq:T4}
\end{eqnarray}
where $\gamma_{m}(R)=\dd^{m}\log \sigma^{2}(R)/\dd(\log R)^{m}$. 
Let us also recall the result for the reduced cumulants $S_n=\langle\rho^n\rangle_c/\sigma_\rho^{2(n-1)}$ of the top-hat filtered density field
\begin{eqnarray}
S_{3}&=&\frac {34}{7}+\gamma_{1}(R),\label{eq:S3}\\
S_{4}&=&\frac{60712}{1323} + \frac{62}{3}\gamma_{1}(R)+ \frac{7}{3} \gamma_{1}(R)^2 + 
\frac 2 3 \gamma_{2}(R) \label{eq:S4}\,.
\end{eqnarray}

\begin{figure}
\includegraphics[width=0.98\columnwidth]{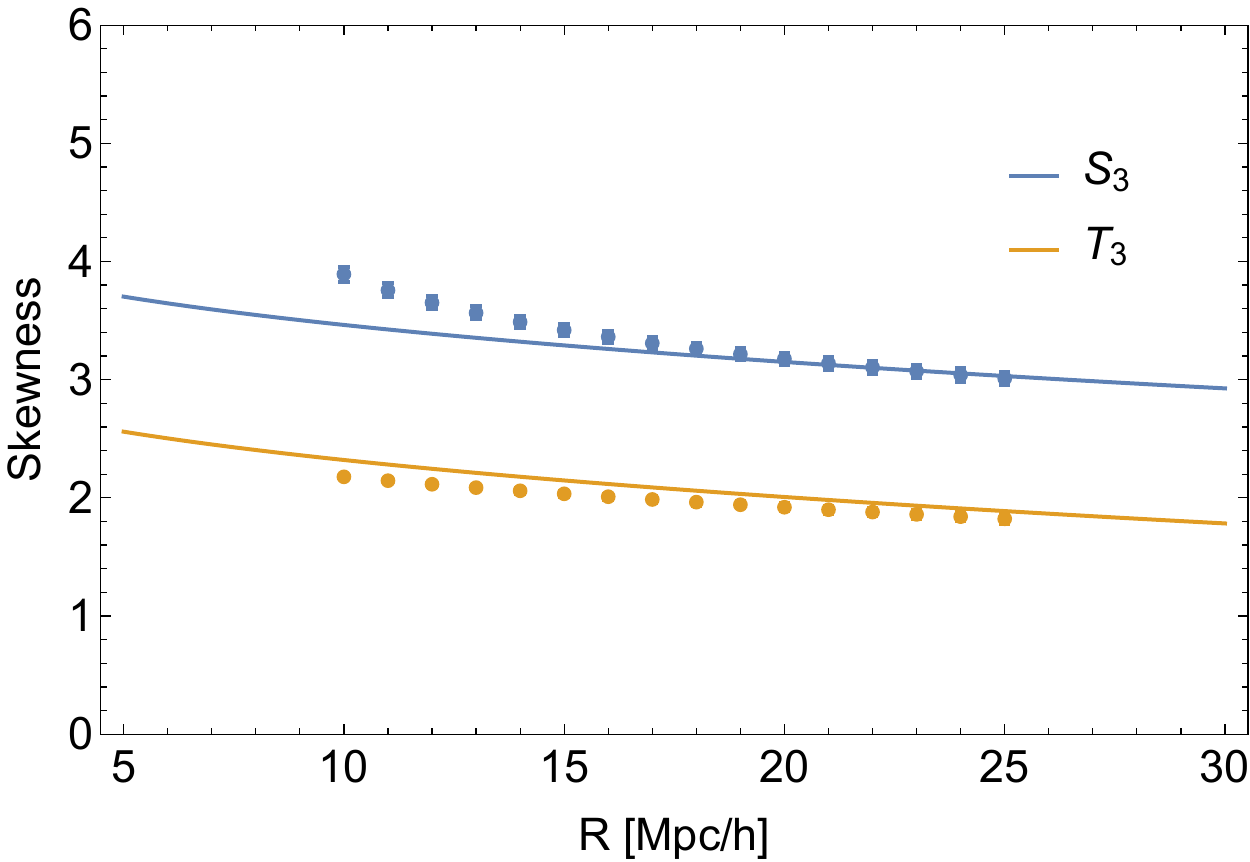}\\
\includegraphics[width=0.98\columnwidth]{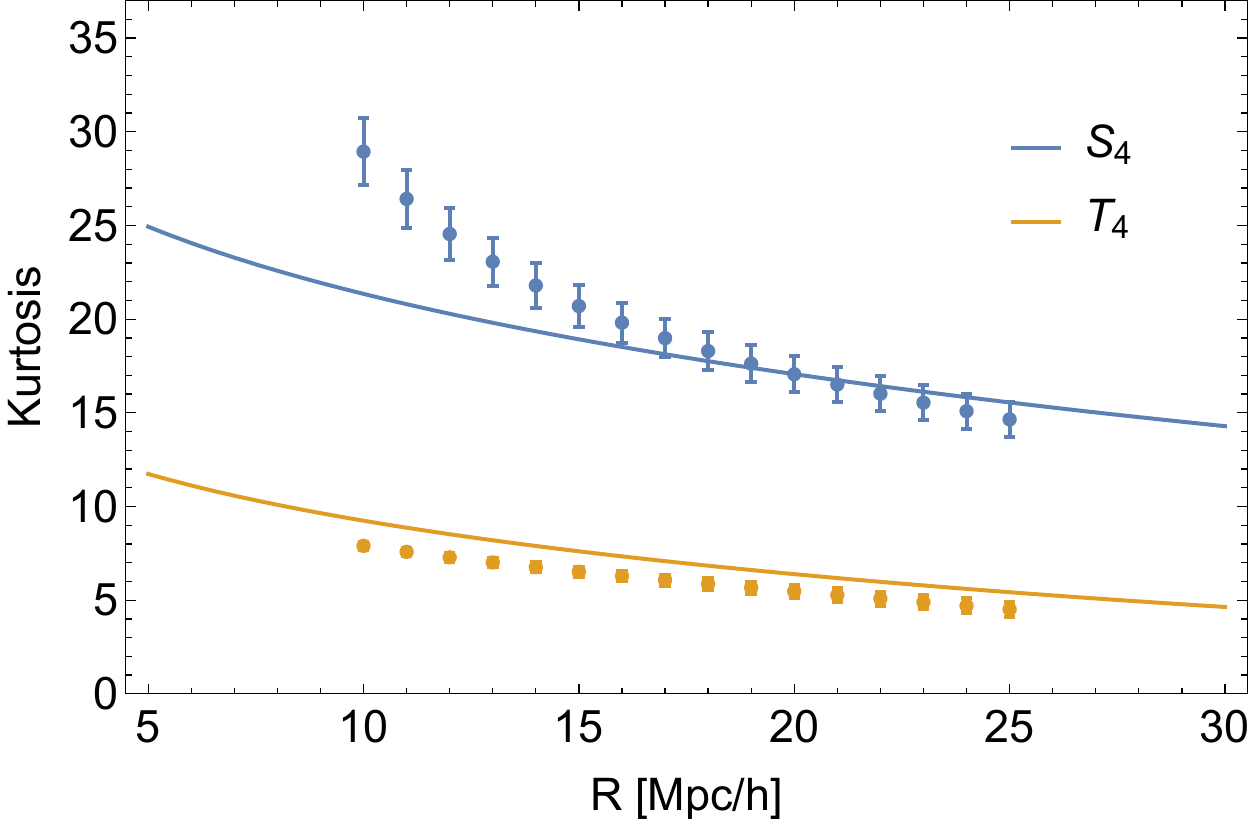}
   \caption{The first two reduced cumulants being the skewness (upper panel) and kurtosis (lower panel) of the density, $S_{3|4}$, and velocity divergence, $T_{3|4}$ measured in the simulations (points error bars) compared to their tree-order prediction (solid lines) for different radii, $R$, of the top-hat filter.}
\label{fig:cumulants}
\end{figure}

The measured skewness and kurtosis of the density and velocity divergence fields are compared to their tree-order PT prediction in Fig.~\ref{fig:cumulants} which improves upon early results by \cite{BernardeauVdW95}.
   
   \begin{figure}
\centering
\includegraphics[width=0.98\columnwidth]{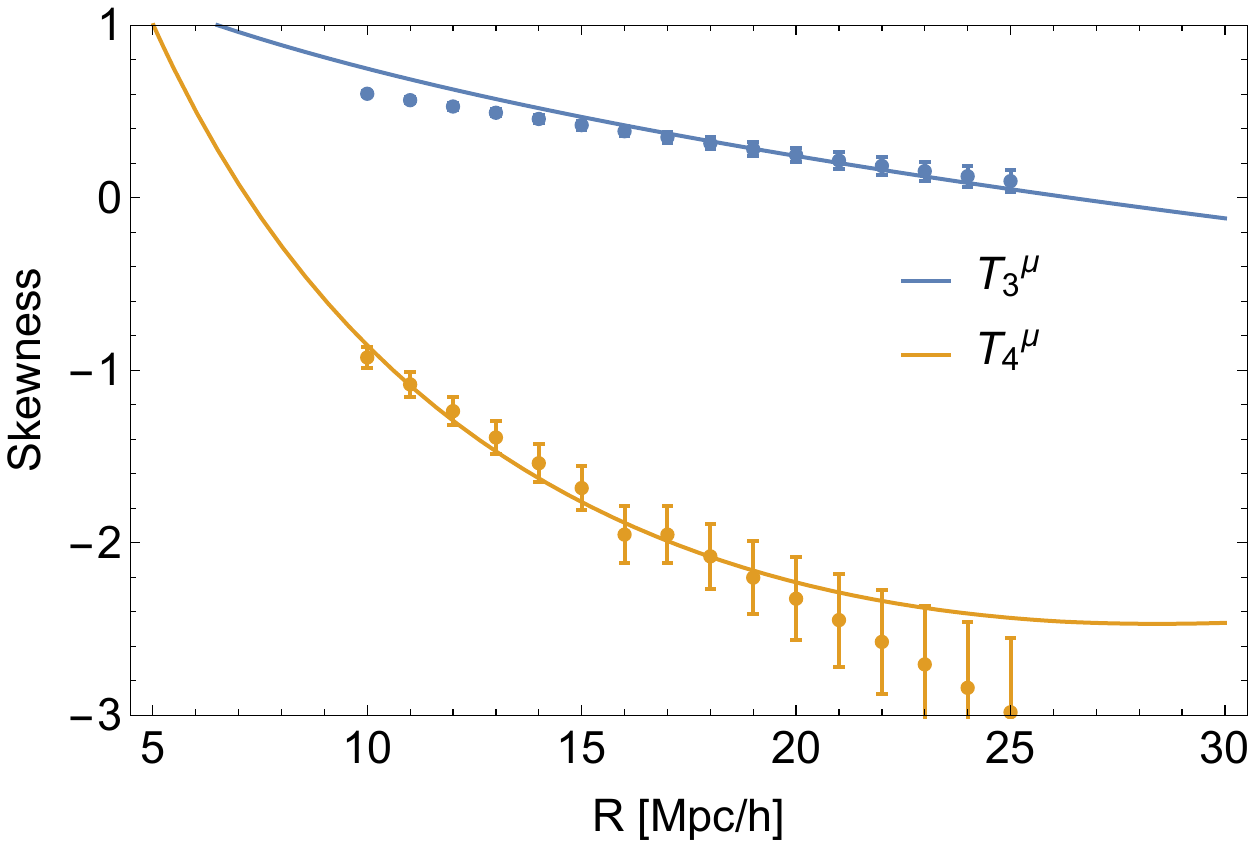}
   \caption{
Same as Fig.~\ref{fig:cumulants}  for the skewness and kurtosis of the log-variable associated to the velocity divergence $\log (\nu+\tilde\theta)$.
\label{fig:cumulants-log}
   }
   \end{figure}

In this paper, we also consider the log-variable $\mu=\log \nu+\tilde\theta$ (or equivalently $\mu=\log (1+\tilde\theta/\nu)$ as they only differ by their mean) whose first cumulants  follow
\begin{eqnarray}
T_{3}^{\mu}&=&\nu T_{3}-3+{\cal O}(\sigma^{2}),\\
T_{4}^{\mu}&=&\nu^{2}T_{4}-12\nu T_{3}+20+{\cal O}(\sigma^{2})\,,
\end{eqnarray}
in complete analogy to the  cumulants of the log-density $\mu=\log \rho$
\begin{eqnarray}
S_{3}^{\mu}&=&S_{3}-3+{\cal O}(\sigma^{2}),\\
S_{4}^{\mu}&=&S_{4}-12 S_{3}+20+{\cal O}(\sigma^{2})\,,
\end{eqnarray}
once the cumulants of $\rho$ are replaced by the cumulants of $\tilde\theta/\nu$ which in turn can be expressed as a function of the cumulants of $\tilde\theta$ and $\nu$.
It has to be emphasised here that according to the approximation given by equation~(\ref{eq:sc-veldiv}), $\rho\approx\nu\log (1+\tilde\theta/\nu)$ so that we have the following approximate relations
\begin{align}
T_{3}^{\mu}\approx \nu S_{3}^{\mu},\\
T_{4}^{\mu}\approx \nu^{2} S_{4}^{\mu},
\end{align}
which is true at tree order for the skewness (as expected since the $\nu$-approximation we use for the spherical collapse is designed to reproduce the zero variance limit of the skewness) but not the kurtosis.
Despite those approximate relations, we stick to the variable  $\mu=\log (1+\tilde\theta/\nu)$ instead of $\nu\mu$ for aesthetic purposes\footnote{We note here that this simple link between the log density and the log variable associated to the velocity divergence implies that applying the large-deviation principle to $\log \rho$ or $\mu=\log (1+\tilde\theta/\nu)$ is equivalent in the $\nu$-approximation we use in this paper.}.

We compare the above-defined tree-order PT predictions of $\log \nu+\tilde\theta$ to measurements in our simulation in Fig.~\ref{fig:cumulants-log}. Similar to what was found in \cite{Uhlemann16} for the log density, the log-variable associated to the velocity divergence, $\mu=\log \nu+\tilde\theta$, have cumulants that are closer to the tree-order predictions. This motivates the choice to rely on a large-deviation principle for the log variable instead of the velocity divergence itself. In addition, we showed in the main text that it allows for fully analytical predictions for the PDF.

Note also that a perturbative expansion can be used to map the variance of the velocity divergence to the variance of the density thanks to the approximation of the spherical collapse dynamics, equation~(\ref{eq:density-to-veldif}). It reads
\begin{equation}
\sigma_{\tilde\theta}^{2}=\sigma_{\rho}^{2}+\sigma_{\rho}^{4}\left(\frac{88}{147}-\frac{8S_{3}}{21}\right).
\end{equation}
In practice, this perturbative relation can be  improved in the quasi-linear regime by using the fully non-linear relation given by equation~\eqref{eq:saddlePDFrhothetaN-cell} at the level of the PDFs.

Beyond one-cell statistics, the large deviation principle can be used to also compute mixed cumulants involving the field (density and velocity divergence) at different scales. To do so, let us first write down the exact spherical collapse mapping as
\begin{align}
\rho_{\rm SC}(\tau)=\sum_{i\geq0}\frac{\nu_{i}}{i!}\tau^{i},\label{eq:nuSC}\\
\tilde\theta_{\rm SC}(\tau)=\sum_{i\geq1}\frac{\mu_{i}}{i!}\tau^{i},\label{eq:muSC}\
\end{align}
with $\nu_{i}$ and $\mu_{i}$ the usual spherically averaged PT kernels \citep{2002PhR...367....1B}
\begin{align}
\nu_{i}=i!\int \dd\Omega_{1}\dots\dd\Omega_{i}F_{i}({\bf k_{1}},\dots,{\bf k_{i}}),\\
\mu_{i}=i!\int \dd\Omega_{1}\dots\dd\Omega_{i}G_{i}({\bf k_{1}},\dots,{\bf k_{i}}).
\end{align}
In particular, one can  check that in an Einstein-de Sitter Universe, $\nu_{1}=1$, $\nu_{2}=34/21$, $\nu_{3}=682/189$, $\nu_{4}=446440/43659$ and for the velocity divergence $\mu_{1}=1$, $\mu_{2}=26/21$, $\mu_{3}=142/63$, $\mu_{4}=236872/43659$.

From equations~(\ref{eq:nuSC}-\ref{eq:muSC}), one can  compute the rate function of $\phi=\rho$ or $\tilde\theta$ and by Legendre transform the corresponding one-cell cumulant generating function $\varphi(\lambda)$. In particular, one can  check that the one-cell cumulants can be computed as
\begin{equation}
\left\langle \phi^{2+p} \right\rangle_{c}=\left.\frac{\partial^{2+p} \varphi}{\partial \lambda^{2+p}}(\lambda=0)=\left[\frac{1}{\Psi''(\phi)}\frac{\dd}{\dd\phi}\right]^{p}\left(\frac{1}{\Psi''(\phi)}\right)\right\vert_{\tau=0}\nonumber
\end{equation}
for all $p\geq0$. Note that the condition $\tau=0$ will translate into either $\rho=1$ or $\tilde\theta=0$. This procedure allows to compute  the successive cumulants of the density and the velocity divergence as a function of the spectral parameters $\gamma_{p}$ and the spherical collapse parameters $\nu_{i}$ and $\mu_{i}$. In particular, it allows to recover the value of the skewness and kurtosis quoted in equations~(\ref{eq:T3}-\ref{eq:S4}) and to compute all higher order cumulants.

The same procedure can be applied for multiple concentric cells in order to compute the mixed cumulants of densities and velocity divergences at different scales.
Again relying on equations~(\ref{eq:nuSC}-\ref{eq:muSC}), the two-cell rate function and cumulant generating function $\varphi(\lambda_{1},\lambda_{2})$ can be  computed. The two-cell cumulants then follow
\begin{align}
\left\langle \phi_{1}^{1+p}\phi_{2}^{1+q} \right\rangle_{c}=\frac{\partial^{2+p} \varphi}{\partial \lambda^{2+p}}(\lambda_{1}=0,\lambda_{2}=0)\nonumber\\
=D_{1}^{p}D_{2}^{q}\left.\left(\frac{-\Psi_{12}}{\det\Psi_{ij}}\right)\right\vert_{\tau_{i}=0}\,,
\end{align}
where  the operators $D_{1}$ and $D_{2}$  are defined as follows
\begin{align}
D_{1}=\left[\frac{\Psi_{22}}{\det\Psi_{ij}}\frac{\partial}{\partial\phi_{1}}-\frac{\Psi_{12}}{\det\Psi_{ij}}\frac{\partial}{\partial\phi_{2}}\right],\\
D_{2}=\left[\frac{-\Psi_{12}}{\det\Psi_{ij}}\frac{\partial}{\partial\phi_{1}}+\frac{\Psi_{11}}{\det\Psi_{ij}}\frac{\partial}{\partial\phi_{2}}\right].
\end{align}
In particular, following this approach, \cite{Bernardeau14} showed that 
\begin{eqnarray}
\langle \rho_{1}^{2}\rho_{2}\rangle_{c}\!\!\!\!\!\!&=&\!\!\!\!\!\!
\nu _2 \sigma^2({R_1,R_2}) \left(\sigma^2({R_1, R_2})+2 \sigma^2({R_1,R_{1}})\right)\nonumber\\
&+&\!\!\!\!\!\!\frac{2}{3}  \sigma^2({R_1},R_{1}) \frac{R_{1}\partial}{\partial R_1}\sigma^2({R_1,R_2)}\nonumber\\
&+&\!\!\!\!\!\!\frac{1}{3}\sigma^2({R_1,R_2}) \left(2  \frac{R_2\partial}{\partial R_2}\sigma^2({R_1,R_2})\!+\!\frac{R_{1}\dd}{ \dd R_1}\sigma^2({R_1},R_{1})\!\right)\!.\nonumber
\end{eqnarray}
A similar result can now be obtained for the mixed cumulants of the velocity divergence field
\begin{eqnarray}
\langle \tilde\theta_{1}^{2}\tilde\theta_{2}\rangle_{c}\!\!\!\!\!\!&=&\!\!\!\!\!\!
\mu _2 \sigma^2({R_1,R_2}) \left(\sigma^2({R_1, R_2})+2 \sigma^2({R_1,R_{1}})\right)\nonumber\\
&+&\!\!\!\!\!\!\frac{2}{3}  \sigma^2({R_1},R_{1}) \frac{R_{1}\partial}{\partial R_1}\sigma^2({R_1,R_2)}\nonumber\\
&+&\!\!\!\!\!\!\frac{1}{3}\sigma^2({R_1,R_2}) \left(2  \frac{R_2\partial}{\partial R_2}\sigma^2({R_1,R_2})\!+\!\frac{R_{1}\dd}{ \dd R_1}\sigma^2({R_1},R_{1})\!\right)\!,\nonumber
\end{eqnarray}
and for the mixed cumulants of the velocity divergence and density fields
\begin{eqnarray}
\langle \rho_{1}\tilde\theta_{2}^{2}\rangle_{c}\!\!\!\!\!\!&=&\!\!\!\!\!\!
\sigma^2({R_1,R_2}) \left(\nu_{2}\sigma^2({R_1, R_2})+2\mu _2  \sigma^2({R_2,R_{2}})\right)\nonumber\\
&+&\!\!\!\!\!\!\frac{2}{3}  \sigma^2({R_2},R_{2}) \frac{R_{2}\partial}{\partial R_2}\sigma^2({R_1,R_2)}\nonumber\\
&+&\!\!\!\!\!\!\frac{1}{3}\sigma^2({R_1,R_2}) \left(2  \frac{R_2\partial}{\partial R_2}\sigma^2({R_1,R_2})\!+\!\frac{R_{1}\dd}{ \dd R_1}\sigma^2({R_1},R_{1})\!\right)\!,\nonumber\\
\langle \rho_{1}^{2}\tilde\theta_{2}\rangle_{c}\!\!\!\!\!\!&=&\!\!\!\!\!\!
\sigma^2({R_1,R_2}) \left(\mu_{2}\sigma^2({R_1, R_2})+2\nu _2  \sigma^2({R_1,R_{1}})\right)\nonumber\\
&+&\!\!\!\!\!\!\frac{2}{3}  \sigma^2({R_1},R_{1}) \frac{R_{1}\partial}{\partial R_1}\sigma^2({R_1,R_2)}\nonumber\\
&+&\!\!\!\!\!\!\frac{1}{3}\sigma^2({R_1,R_2}) \left(2  \frac{R_1\partial}{\partial R_1}\sigma^2({R_1,R_2})\!+\!\frac{R_{2}\dd}{ \dd R_2}\sigma^2({R_2},R_{2})\!\right)\!.\nonumber
\end{eqnarray}
    \begin{figure}
\centering
\includegraphics[width=0.98\columnwidth]{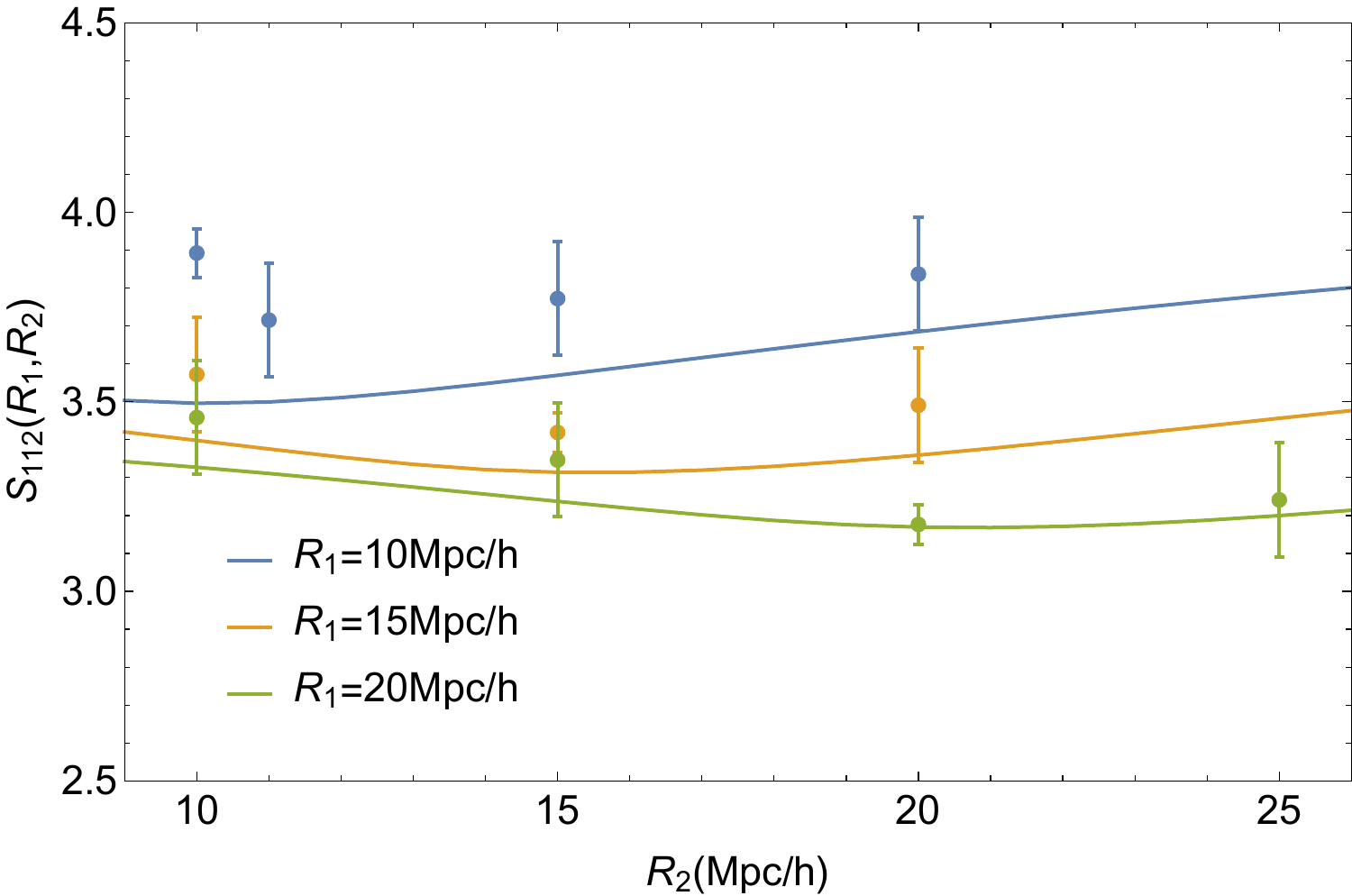}
   \caption{
Mixed third order cumulants of the density field as defined in equation~(\ref{eq:S112}) for $R_{1}$ and $R_{2}$ as labeled. The tree-order prediction is shown with solid lines and the measurements with error bars. Note that in this plot, we estimate the cumulants composed of two different scales from the measured PDF and we display fixed error bars of absolute size 0.15 as this is the typical error we make when using the PDF to estimate the skewness of the density rather than direct measurements in the simulation.
\label{fig:S112}
   }
   \end{figure}
      \begin{figure}
\centering
\includegraphics[width=0.98\columnwidth]{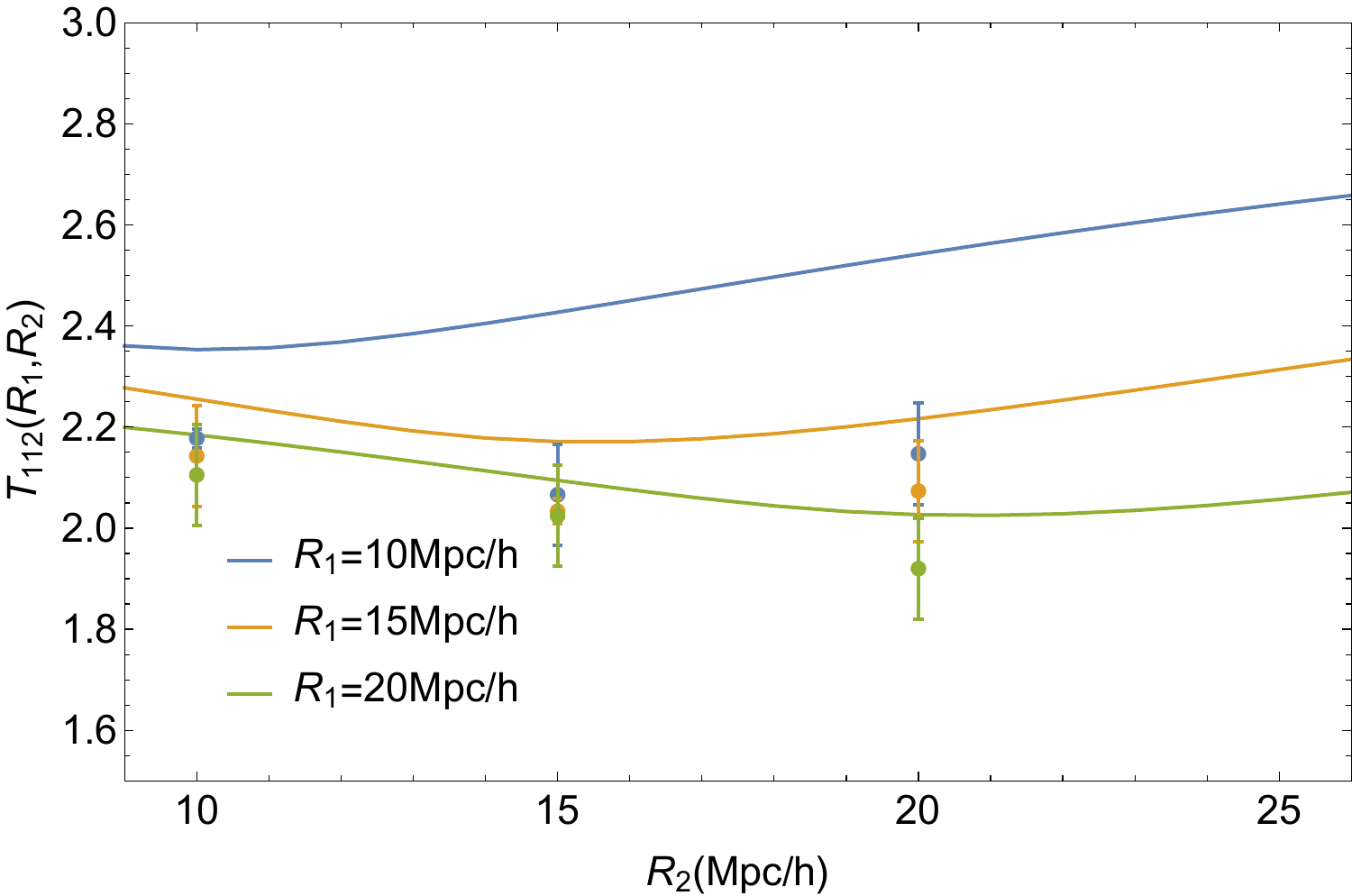}
   \caption{
Same as Fig.~\ref{fig:S112} for the two-cell skewness of the velocity divergence. Here we assume error bars are of the order of 0.1 given the error we make in estimating the skewness from the PDF rather than from direct measurements.
\label{fig:T112}
   }
   \end{figure}
   The above formulae allow us to compute the corresponding multiscale  reduced higher order moments. For instance,
 Fig.~\ref{fig:S112} compares measurements and tree-order predictions of the two-scale skewness of the density field defined as
 \begin{equation}
 \label{eq:S112}
 S_{112}=\frac{3\langle \rho_{1}^{2}\rho_{2}\rangle_{c}}{\sigma^2({R_1,R_2}) \left(\sigma^2({R_1, R_2})+2 \sigma^2({R_1,R_{1}})\right)},
 \end{equation}
 which is exactly the skewness, $S_{3}$, when $R_{1}=R_{2}$. For radii $R_{1},R_{2}\gtrsim15$Mpc$/h$ at redshift 0, we find that measurements agree very well with the tree-order prediction. Below this scale, some departure starts to appear as non-linear corrections arise. Note that tree order predictions are computed by integrating the linear power spectrum such that
 \begin{align}
 \sigma^2({R_1, R_2})=\int\frac{\dd^{3} k}{(2\pi)^{3}}P^{\rm lin}(k)W_{\rm 3D}(k R_{1})W_{\rm 3D}(k R_{2}),\nonumber\\
  R_{1}\frac{\partial}{\partial R_1}\sigma^2({R_1,R_2)}=\int\frac{\dd^{3} k}{(2\pi)^{3}}P^{\rm lin}(k)\tilde W_{\rm 3D}(k R_{1}) W_{\rm 3D}(k R_{2})\nonumber,\\
 R_{2}\frac{\partial}{\partial R_2}\sigma^2({R_1,R_2)}=\int\frac{\dd^{3} k}{(2\pi)^{3}}P^{\rm lin}(k)W_{\rm 3D}(k R_{1})\tilde W_{\rm 3D}(k R_{2}),\nonumber
 \end{align}
 where the logarithmic derivative of the top-hat filter reads
 \begin{equation}
 \tilde W_{\rm 3D}(x)=-3\left(W_{\rm 3D}(x)+\frac{\sin x}{x}\right)\,.
 \end{equation}
 The same comparison as in Fig.~\ref{fig:S112} is also performed for the velocity divergence field. The result is displayed in Fig.~\ref{fig:T112}. The convergence towards the tree-order prediction seems slower for the velocity divergence than the density as we found consistent results for $R_{1},R_{2}\gtrsim20$Mpc$/h$ at redshift 0
 in this case. This might come from the fact that velocities are more sensitive to previrialisation motions and therefore depart earlier from the tree-order expectations. 
 
\begin{figure}
\centering
\includegraphics[width=0.98\columnwidth]{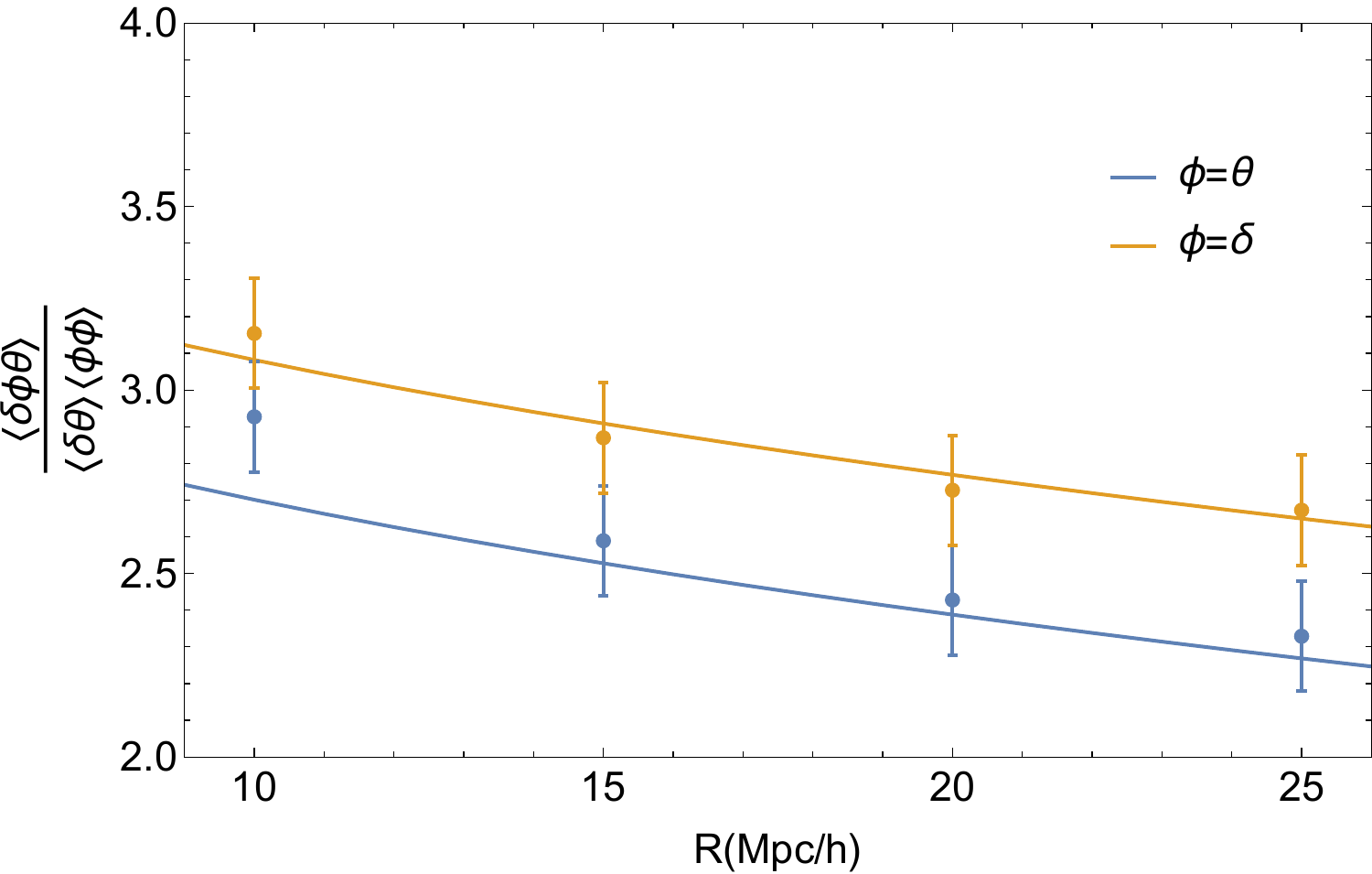}
   \caption{
Same as Fig.~\ref{fig:S112} for the mixed skewness of the velocity divergence and density at the same scale $R$. Here we assume error bars are of the order of 0.15.
\label{fig:dtt}
   }
   \end{figure}

One can also compute the joint cumulants of velocities and densities in order to better assess the scatter in the density-velocity relation. As an illustration we measure the third order cumulants involving densities and velocities at the same scale  and we compare them in Fig.~\ref{fig:dtt} to their respective tree-order prediction
\begin{align}
\label{eq:dtt}
\frac{\left\langle\rho\tilde\theta\tilde\theta\right\rangle_{c}}{\left\langle\delta\tilde\theta\right\rangle_{c}\left\langle\tilde\theta\tilde\theta\right\rangle_{c}}=\nu_{2}+2\mu_{2}+\gamma(R)+{\cal O}(\sigma^{2}),
\\\label{eq:ddt}
\frac{\left\langle\rho\rho\tilde\theta\right\rangle_{c}}{\left\langle\delta\tilde\theta\right\rangle_{c}\left\langle\rho\rho\right\rangle_{c}}=2\nu_{2}+\mu_{2}+\gamma(R)+{\cal O}(\sigma^{2}).
\end{align}
Those mixed cumulants are very accurately fitted by their tree-order prediction (provided we divide by the right combination of variances in equations~(\ref{eq:dtt}-\ref{eq:ddt})). 

All higher order mixed cumulants (for instance the two-scale kurtosis) can be obtained  following the same scheme as used here for the two-scale skewness. 
Given the length of those formula, we do not show them in this paper.

\end{document}